\documentclass[aps,prb,reprint,showpacs,superscriptaddress]{revtex4-1}

\usepackage{amsmath,amssymb,times,bm}
\usepackage[pdftex]{graphicx,color}
\usepackage[pdftex,bookmarks=false,pdfborder={0 0 0},colorlinks=true,linkcolor=magenta,citecolor=cyan,urlcolor=cyan,raiselinks]{hyperref}
\usepackage[all]{hypcap}
\graphicspath{{fig2/},{./}}
\makeatletter
\newcommand*{\balancecolsandclearpage}{%
  \close@column@grid
  \clearpage
  \twocolumngrid
}

\newcounter{smsection}
\newcommand{\smsection}[1]{
	\refstepcounter{smsection}
	\vspace{2ex}
	\begin{center}
	\textbf{
	\@currentlabel. \, #1
	}
	\end{center}
	\par
}
\newcounter{smfig}
\makeatother

\begin{document}
%%%%%%%%%%
% information
%%%%%%%%%%
\title{Two-dimensional $p$-wave superconducting states with magnetic moments \\ 
 on a conventional $s$-wave superconductor}
\date{\today}
\author{Sho Nakosai}
\affiliation{Department of Applied Physics, University of Tokyo, Tokyo 113-8656, Japan}
\author{Yukio Tanaka}
\affiliation{Department of Applied Physics, Nagoya University, Aichi 464-8603, Japan}
\author{Naoto Nagaosa}
\affiliation{RIKEN Center for Emergent Matter Science (CEMS), Wako 351-0198, Japan}
\affiliation{Department of Applied Physics, University of Tokyo, Tokyo 113-8656, Japan}

\begin{abstract}
Unconventional superconductivity induced by the magnetic moments in a conventional spin-singlet $s$-wave 
superconductor is theoretically studied. By choosing the spin directions of these moments, one can design 
spinless pairing states appearing within the $s$-wave superconducting energy gap. 
It is found that the helix spins produce $p_{x}+p_{y}$-wave state while the skyrmion crystal configuration 
$p_{x}+ip_{y}$-wave like state. 
Nodes in the energy gap and the zero energy flat band of Majorana edge states exist in the former one, 
while the chiral Majorana channels along edges of the sample and the zero energy Majorana bound 
state at the core of the vortex appear in the latter case. 
\end{abstract}
\maketitle

%===========================================================
% INSRODUCTION
%===========================================================
\textit{Introduction.---}
Unconventional superconducting states are one of the most important issues in current condensed matter 
physics.~\cite{RevModPhys.63.239, PhysRevLett.43.1892, JPSJ.81.011013} 
Although most of the superconductors show the conventional spin-singlet $s$-wave pairing, strongly correlated 
materials 
sometimes show unconventional pairing since the on-site pairing is suppressed by the repulsive interaction. 
However, the discovery of the unconventional pairings relies on serendipity to some degree, and their 
theoretical designs and artificial fabrications are highly desired. Especially, recent intensive interest in 
the topological superconductivity and consequent Majorana fermions enhance the importance of this topics 
since Majorana fermions are the leading candidate for the platform of the quantum 
computation.~\cite{PhysRevB.61.10267, PhysRevLett.86.268, NuclPhysB.360.362, RevModPhys.80.1083, 
RevModPhys.83.1057, RepProgPhys.75.076501}

A promising proposal for realization of a topological superconducting state is the combined system of 
semiconducting nanowire 
with $s$-wave superconductor under an external magnetic field. The spin-orbit interaction and the magnetic 
field reduce the degrees of freedom of electrons concerning superconducting states, and effectively 
generate spinless $p$-wave superconductor.~\cite{PhysRevLett.104.040502, PhysRevB.81.125318, 
PhysRevLett.105.077001, PhysRevLett.105.177002, JPhysCondMatt.25.233201} 
As for one-dimensional system, signals suggesting Majorana fermions at the 
ends of the nanowire have been observed in some experimental setups.~\cite{Science.336.1003, NanoLett.12.6414, 
NatPhys.8.795, NatPhys.8.887} 
There are other routes for creating topological superconductors; spin-singlet superconductor 
deposited on the topological insulator.~\cite{PhysRevLett.100.096407, PhysRevLett.104.067001} 
superfluid of cold atoms with 
laser-generated effective spin-orbit interaction.~\cite{PhysRevLett.103.020401}, 
aligned quantum dots connected by $s$-wave superconductors,~\cite{NatCommun.3.964} and magnetic moments
in $s$-wave superconductors~\cite{PhysRevB.82.045127, PhysRevB.84.195442, PhysRevB.85.144505, PhysRevB.88.020407} 
or nodal superconductors.~\cite{PhysRevLett.110.096403} 
The last ones are significantly distinct in that they don't explicitly require spin-orbit interaction in the 
system. With respect to the cooperation between magnetic moment and superconductivity, it has been known that 
the bound states are created 
around the impurities with the energy inside the bulk superconducting gap (not necessarily zero 
energy).~\cite{ProgTheorPhys.40.435, SovPhysJETP.29.1101, ActaPhysSin.21.75} 
The modulation of the local density of 
states by a single magnetic impurity has been observed in the experiment.~\cite{Science.275.1767} 
The authors of Refs~\onlinecite{PhysRevB.84.195442, PhysRevB.88.020407} considered the one-dimensional array 
of the magnetic impurities, and studied the possibility of Kitaev state with the Majorana bound states at the 
ends of the array. 
The influence of magnetic moments on a superconductor by the proximity effect has been intensively 
studied albeit in different interests.~\cite{RevModPhys.77.935, RevModPhys.77.1321, PhysRevB.79.054523} 
On the other hand, it has been recognized that the spin-orbit interaction at the interface results in Rashba 
type interaction and hence non-collinear spin configuration is organized.~\cite{NatNano.8.152} 
Especially, the skyrmion crystal state is observed at the interface of Fe and Ir.~\cite{NatPhys.7.713} 
Therefore, the magnetic proximity effect of non-collinear moments to a superconductor becomes a realistic and 
important issue. 
\begin{figure}[b]
\centering
\includegraphics[bb=0 0 1441 508, width=\hsize]{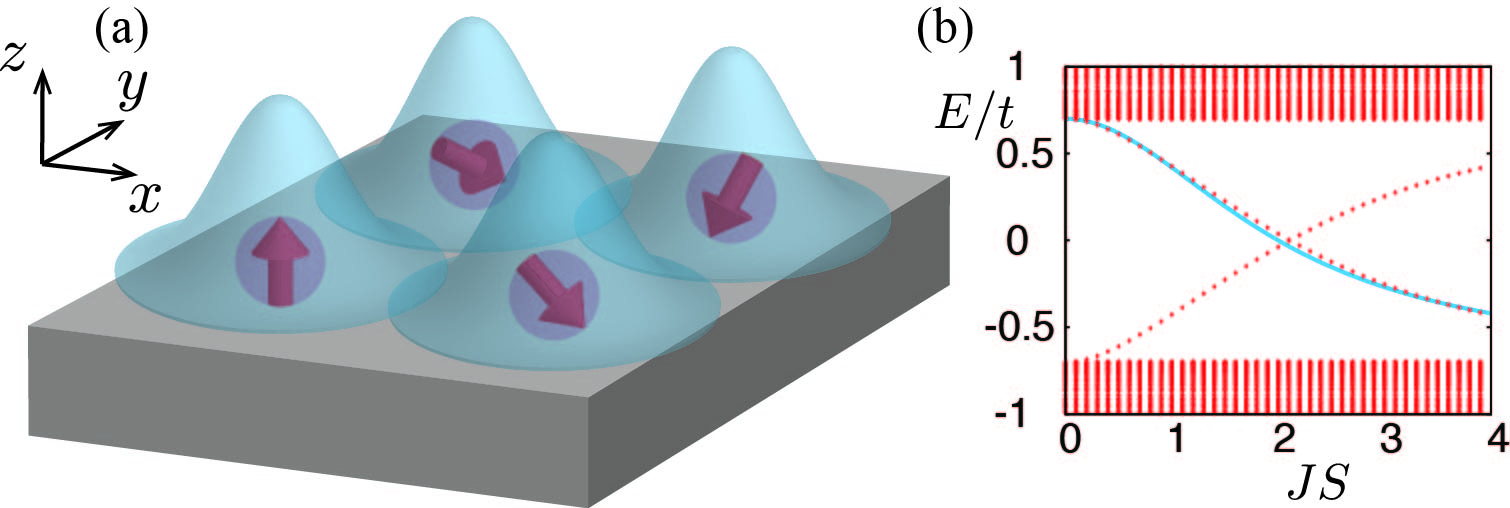}%
\caption{\label{fig:idea}
	(a) Schematic illustration for the formation of an effective lattice model from the bound states localized 
	around magnetic moments on the surface of $s$-wave superconductor (see Eq.~(\ref{eq:wf})). 
	(b) Energy levels of quasiparticles obtained by a tight binding model calculation with a single moment with 
	$\Delta_{0}/t=0.7$ in Eq.~(\ref{eq:tb}). $J$ is the coupling constant between electrons and 
	magnetic moment and $S$ is the magnitude of the spin moment. The solid (blue) curve shows the analytical 
	solution $E_{0}$ for the continuum model (see the main text).
	}
\end{figure}

In this paper, we propose a generic principle to design unconventional superconductivity in terms of 
non-collinear/non-coplanar configurations of magnetic moments on the surface of an $s$-wave superconductor.
We derive an effective model constituted from the bound states around magnetic moments. The effective pair 
potentials as well as transfer integrals in the effective model depend on the directions of two neighboring moments. 
We show that a $p_{x}+p_{y}$-wave pairing state with nodes in the energy gap is generated by a non-collinear 
helical spin configuration, and moreover, we design a topological $p_{x}+ip_{y}$-wave like state by means of 
a non-coplanar skyrmion crystal configuration of moments, as evidenced by chiral Majorana channels along the 
edges of the system and zero energy Majorana bound states at the cores of vortices. 

%===========================================================
% MODEL
%===========================================================
\textit{Model.---}
Figure~\ref{fig:idea} (a) shows a schematic illustration of the present model. 
We analyze the following tight-binding Hamiltonian describing double exchange model with the superconducting order 
parameter defined on a square lattice 
\begin{align}
\label{eq:tb}
H = 
& - \sum_{\langle ij \rangle\sigma} t c_{i\sigma}^{\dagger} c_{j\sigma} 
- \sum_{i} \mu c_{i\sigma}^{\dagger} c_{i\sigma} 
\nonumber \\
& + \sum_{i} \Delta_0 \left( c_{i\uparrow}^{\dagger} c_{i\downarrow}^{\dagger} + \mathrm{H.c.} \right)
- \sum_{i} J \bm{S}_{i}\cdot\bm{\sigma}_{\alpha\beta}c_{i\alpha}^{\dagger} c_{i\beta}. 
\end{align}
The first three terms describe a conventional spin-singlet $s$-wave superconductor with the transfer integral 
$t$, the chemical potential $\mu$, and the pairing potential $\Delta_{0}$. In addition, electrons couple 
with magnetic moments located at sites $i$'s with the strength $J$ through double exchange mechanism.  
This model can describe the interface between a bulk $s$-wave superconductor and a magnetic material. 
We assume 
that the pairing potential is not affected by magnetic moments, which are supposed to be solidly ordered. 
Below we construct unconventional superconducting states with some particular structures of magnetic moments. 
We derive an effective model in the aim of choosing appropriate magnetic order for intended unconventional states 
before directly solving Eq.~(\ref{eq:tb}). 
First, we start with the case of a single moment in a superconductor. The Bogoliubov-de Gennes equations are 
given by 
\begin{align}
\label{eq:bdg}
	(\xi_k - E) u_{k\uparrow} - \frac{JS}{V} \sum_l u_{l\uparrow} + \Delta_0 v_{k\downarrow} &= 0, \nonumber \\
	(\xi_k + E) v_{k\downarrow} + \frac{JS}{V} \sum_l v_{l\downarrow} - \Delta_0 u_{k\uparrow} &= 0, 
\end{align}
where we set the origin at the site of moment, $\hbar=1$, and 
$\xi(\bm{k}) = -2t(\cos k_{x} + \cos k_{y}) - \mu $ is the tight-binding dispersion. The numerical result of 
energy level is shown in Fig.~\ref{fig:idea}~(b). The dispersion can be approximated in the continuum limit as 
$\xi(\bm{k}) = \frac{k^2}{2m} - \mu - 4t$ with $m = (2t)^{-1}$. 
We can find solutions with energy $\pm E_{0}=\pm\Delta_0\left[1-(\pi JSN_0/2)^2\right]/
\left[1+(\pi JSN_0/2)^2\right]$, where $N_0$ is the density of states in the normal phase~\cite{ProgTheorPhys.40.435, 
SovPhysJETP.29.1101, ActaPhysSin.21.75, RevModPhys.78.373} (solid line in Fig.~\ref{fig:idea}~(b)). 
With increasing the magnitude of $JS$, $E_0$ changes from $\Delta_0$ to 
$-\Delta_0$ within the bulk superconducting energy gap. The corresponding wave functions are real and 
asymptotically given for $r \to \infty$ as
\begin{align}
\label{eq:wf}
	u_{\uparrow}(\bm{r}) &\sim 
	\frac{\sin(p_{\mathrm{F}}r-\delta_+)}{p_{\mathrm{F}}r} \exp\left[ -\frac{r}{\xi_0} |\sin(\delta_+-\delta_-)|\right],
	\nonumber \\
	v_{\downarrow}(\bm{r}) &\sim 
	\frac{\sin(p_{\mathrm{F}}r-\delta_-)}{p_{\mathrm{F}}r} \exp\left[ -\frac{r}{\xi_0} |\sin(\delta_+-\delta_-)|\right],
\end{align}
where we define some quantities; $\tan\delta_{\pm} = \pm \pi JSN_0 /2$, $p_{\mathrm{F}}$ is the Fermi momentum, 
$v_{\mathrm{F}}=p_{\mathrm{F}}/m$ is the Fermi velocity, and $\xi_{0}=v_{\mathrm{F}}/(\pi\Delta)$.
When the moments are aligned in a lattice, we expect that the bound state around each moment has overlap with 
neighboring bound states. The overlap causes effective transfer integrals and pair potentials among the 
bound states. 
The low energy properties, i.e., in the bulk superconducting gap, can be described by an 
effective BdG lattice model constructed from these bound states. One can find similar arguments in 
Refs.~\onlinecite{PhysRevB.84.195442, PhysRevB.85.144505, PhysRevB.88.020407}.

%===========================================================
% EFFECTIVE MODEL
%===========================================================
\textit{Design of $p$-wave superconducting states.---}
We propose a principle to design the superconducting states appearing within the gap of the host spin-singlet 
$s$-wave superconductor. It will be shown that configurations of magnetic moments play the essential role 
for the emergence of unconventional superconducting states. 
In the last section, we assume the magnetic moment parallel to $+s_{z}$-direction. Here, we introduce a unitary 
transformation for the description of general directions of moments. The coupling term with magnetic moments in 
Eq.~(\ref{eq:tb}) can be transformed as
\begin{align}
\label{eq:ut}
	\psi_{\alpha}^{\dagger} \bm{S} \cdot \bm{\sigma}_{\alpha \beta} \psi_{\beta} =
	(U\psi)^{\dagger} U \bm{S} \cdot \bm{\sigma} U^{\dagger} U \psi =
	\tilde{\psi}^{\dagger} S \sigma_z \tilde{\psi}
\end{align}
by the unitary matrix
\begin{align}
\label{eq:u}
	U^{\dagger} = 
	\begin{pmatrix}
		\cos \frac{\theta}{2} & -e^{-i\phi} \sin \frac{\theta}{2} \\
		e^{i\phi} \sin \frac{\theta}{2} & \cos \frac{\theta}{2}
	\end{pmatrix}
\end{align}
where $\theta$ and $\phi$ are the polar coordinates such that $\bm{S}=S(\sin\theta\cos\phi,\, 
\sin\theta\sin\phi,\, \cos\theta)$. The wave functions for arbitrary spin directions are obtained by 
operating $U^{\dagger}$ on Eqs.~(\ref{eq:wf}). 
Then, the electron operators are expressed for the low energy sector as
\begin{align}
\label{eq:me}
	\psi_{\uparrow} &= 
	\sum_{i} \left[ \cos\frac{\theta_{i}}{2} u_{\uparrow}(\bm{r} - \bm{r}_{i}) \alpha_{i}
		- e^{-i\phi_{i}} \sin\frac{\theta_{i}}{2} v_{\downarrow}^{*}(\bm{r}-\bm{r}_{i}) \alpha_{i}^{\dagger} \right], 
\nonumber \\
	\psi_{\downarrow} &= 
	\sum_{i} \left[ e^{i\phi_{i}} \sin\frac{\theta_{i}}{2} u_{\uparrow}(\bm{r}-\bm{r}_{i}) \alpha_{i}
		+ \cos\frac{\theta_{i}}{2} v_{\downarrow}^{*}(\bm{r} - \bm{r}_{i}) \alpha_{i}^{\dagger} \right],
\end{align}
where $\alpha_{i}$ is the annihilation operator of the bound state around the moment located at site $i$.
By substituting Eqs.~(\ref{eq:me}) into the original Hamiltonian Eq.~(\ref{eq:tb}), we obtain
\begin{align}
\label{eq:eff}
	H_{\mathrm{eff}} = 
	\sum_{i} E_{0} \alpha_{i}^{\dagger} \alpha_{i}
	+ \sum_{\langle ij\rangle} \left[ \bar{t}_{ij}\alpha_{i}^{\dagger} \alpha_{j} 
	+ \left( \bar{\Delta}_{ij} \alpha_{i}^{\dagger} \alpha_{j}^{\dagger} + \mathrm{H.c.} \right) \right],
\end{align}
where $\bar{t}_{ij}$ and $\bar{\Delta}_{ij}$ are effective transfer integrals and pair potentials for the nearest 
neighbor sites $\langle i,j\rangle$ in the present low energy Hamiltonian. We keep them up to the nearest 
neighboring sites. Here, we define
\begin{align}
	\hat{z}_{i} = 
	\begin{pmatrix}
		\cos \frac{\theta_{i}}{2} \\ e^{i\phi_{i}}\sin \frac{\theta_{i}}{2}
	\end{pmatrix},
\end{align}
which represents the spin as $\bm{S}_{i}=S\hat{z}_{i}^{\dagger}\bm{\sigma}\hat{z}_{i}$. 
The effective transfer integrals and pair potentials are represented by
\begin{align}
\label{eq:t}
	\bar{t}_{ij} 
	&= \hat{z_{i}}^{\dagger} \hat{z_{j}} \bar{t}_0, \\
	\label{eq:d}
	\bar{\Delta}_{ij}
	&=  \hat{z_{i}}^{\dagger} i \sigma_y \hat{z_{j}}^{*} \bar{\Delta}_0,
\end{align}
with 
$\bar{t}_0=\int \mathrm{d}\bm{r} \big[ \left( u_{i} \xi(\bm{r}) u_{j}	- v_{i} \xi(\bm{r}) v_{j} \right)+ 
\Delta_0 \left( u_{i} u_{j} + v_{i} v_{j} \right)\big]$, $\bar{\Delta}_0=\int \mathrm{d}\bm{r} \big[\left( 
u_{i} \xi(\bm{r}) v_{j} + v_{i} \xi(\bm{r}) u_{j} \right)+\Delta_0 \left( u_{i} u_{j}	- 
v_{i} v_{j} \right)\big]$, $u_{i}=u_{\uparrow}(\bm{r}-\bm{r}_{i})$, and 
$v_{i}=v_{\downarrow}(\bm{r}-\bm{r}_{i})$. 
Based on these equations, we can design various kinds of superconducting states. 
Note that $\bar{t}_{ij}$ and $\bar{\Delta}_{ij}$ are invariant for the common rotation of both moments at 
sites $i$ and $j$. Namely, these quantities depend only on the relative direction of the two moments. 
The electron spin of the bound state is uniquely determined by the magnetic moment, and hence we have a spinless 
lattice model with controllable parameters depending on the configuration of moments. 

%===========================================================
% NUMERICAL STUDIES
%===========================================================
\textit{Numerical studies.---}
We design two-dimensional superconducting states by choosing appropriate configurations of moments designed from 
Eqs.~(\ref{eq:t}) and (\ref{eq:d}) (see Supplemental Material~\ref{sec:td}). We investigate properties of the 
system \textit{by directly solving the original tight-binding Hamiltonian} (Eq.~(\ref{eq:tb})). 
The calculations are performed with transfer integral $t=1.0$ and on-site superconducting order parameter of 
the host conventional superconductor $\Delta_0=0.7$. 
The magnetic moment is attached to each site. We consider both periodic and open boundary conditions 
to see bulk properties and edge states of the system, respectively. 
Hereafter when we calculate the dispersion of Andreev bound states, we use a cylindrical configuration with 
open boundaries along $y$-direction. 
We consider following two cases; $p_{x} + p_{y}$-wave pairing and $p_{x} + ip_{y}$-wave like pairing.

1. \textit{Nodal superconductor.} 
We can see from Eqs.~(\ref{eq:t}) and (\ref{eq:d}) that $\bar{\Delta}_{ij}$ vanishes when the two neighboring 
moments point the same direction ($\theta_i=\theta_j,\,\phi_i=\phi_j$), while $\bar{t}_{ij}$ vanishes for the 
opposite direction ($\theta_i=\pi-\theta_j,\,\phi_i=\phi_j+\pi$). 
Therefore, non-collinear spin configurations are required to obtain the nontrivial states. 
We know that one-dimensional helical or spiral spin structure generates $p$-wave superconducting 
states.~\cite{PhysRevB.88.020407} 
\begin{figure}
\centering
\includegraphics[width=\hsize]{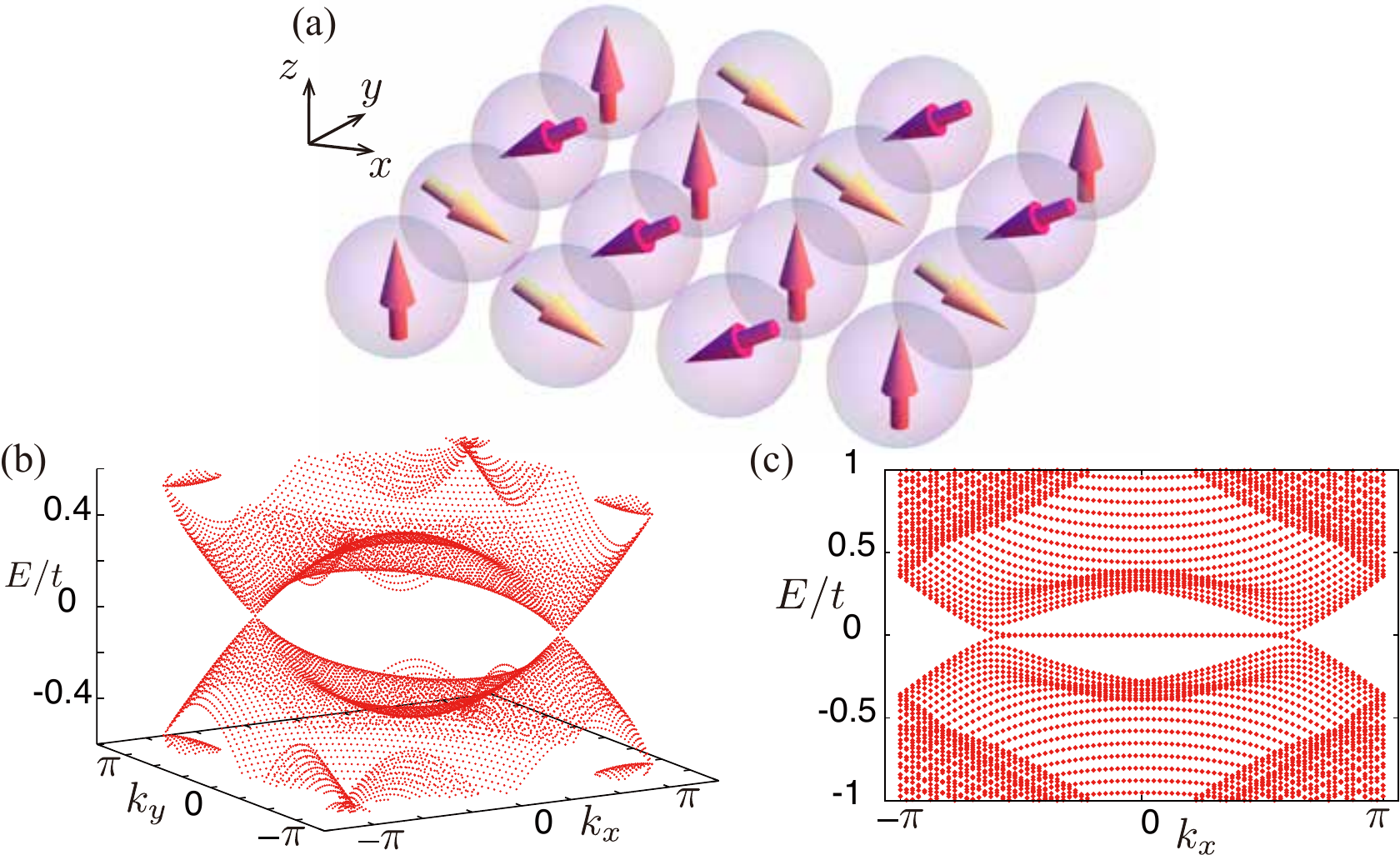}%
\caption{\label{fig:helix}
	(a) Configuration of moments producing effective $p_{x}+p_{y}$-wave pairing. All spins lie in $s_{x}s_{z}$-plane 
		rotating by $2\pi/3$ along both $x$- and $y$-directions, therefore $3\times3$ block constitute a unit cell. 
	(b) Energy spectrum of quasiparticles with the configuration in (a) calculated by tight binding model 
		(Eq.~(\ref{eq:tb})) with $\ t=1.0$ and $\ \Delta_0=0.7$. The spectrum has two point nodes as expected. 
	(c) Energy spectrum with open boundaries. One can see dispersionless Andreev bound states connecting two 
		nodal points. 
	}
\end{figure}
We can generalize this to two-dimensions. We choose the spinor $\hat{z}_{i}$'s so that spins rotate 
by $2\pi/3$ around $s_{y}$-axis along both $x$- and $y$-directions
as shown in Fig.~\ref{fig:helix}~(a) and Fig.~\ref{fig:phelix} in Supplemental Material~\ref{sec:td}; all the moments lie 
in the $s_{x}s_{z}$-plane. 
The resulting state is expected to have real order parameter with $p$-wave pairing for $x$- and $y$-directions.
This can be named as $p_{x}+p_{y}$-wave pairing state.
The present superconducting state is not stabilized as a bulk phase because it is energetically 
disadvantageous compared with chiral $p$-wave pairing without nodal structures. 
However, in the model considered here, induced $p$-wave pairing is localized in the vicinity of the surface of the 
bulk superconductor, and is controlled by the configuration of moments. As a result, $p_{x}+p_{y}$-wave pairing 
state is realized by a single spiral structure of moments. 
We have tested our expectation by numerical calculations solving Eq.~(\ref{eq:tb}). Figure~\ref{fig:helix} (b) shows 
the dispersion of the quasiparticles, where the induced energy gap has nodes on the line of $k_{x} = -k_{y}$. 
We also calculate Andreev bound states at the boundary of the system. 
The resulting dispersions have flat bands, which are believed to be a hallmark of 
unconventional superconductors.~\cite{ProgTheorPhys.76.1237, EurPhysJB.37.349, RepProgPhys.63.1641, 
PhysRevB.83.224511} 
It is noted that flat band of zero energy bound states are realized starting from conventional $s$-wave 
superconducting pairing. One can find in Ref.~\onlinecite{PhysRevB.87.054501} a related work.

\begin{figure}
\centering
\includegraphics[width=\hsize]{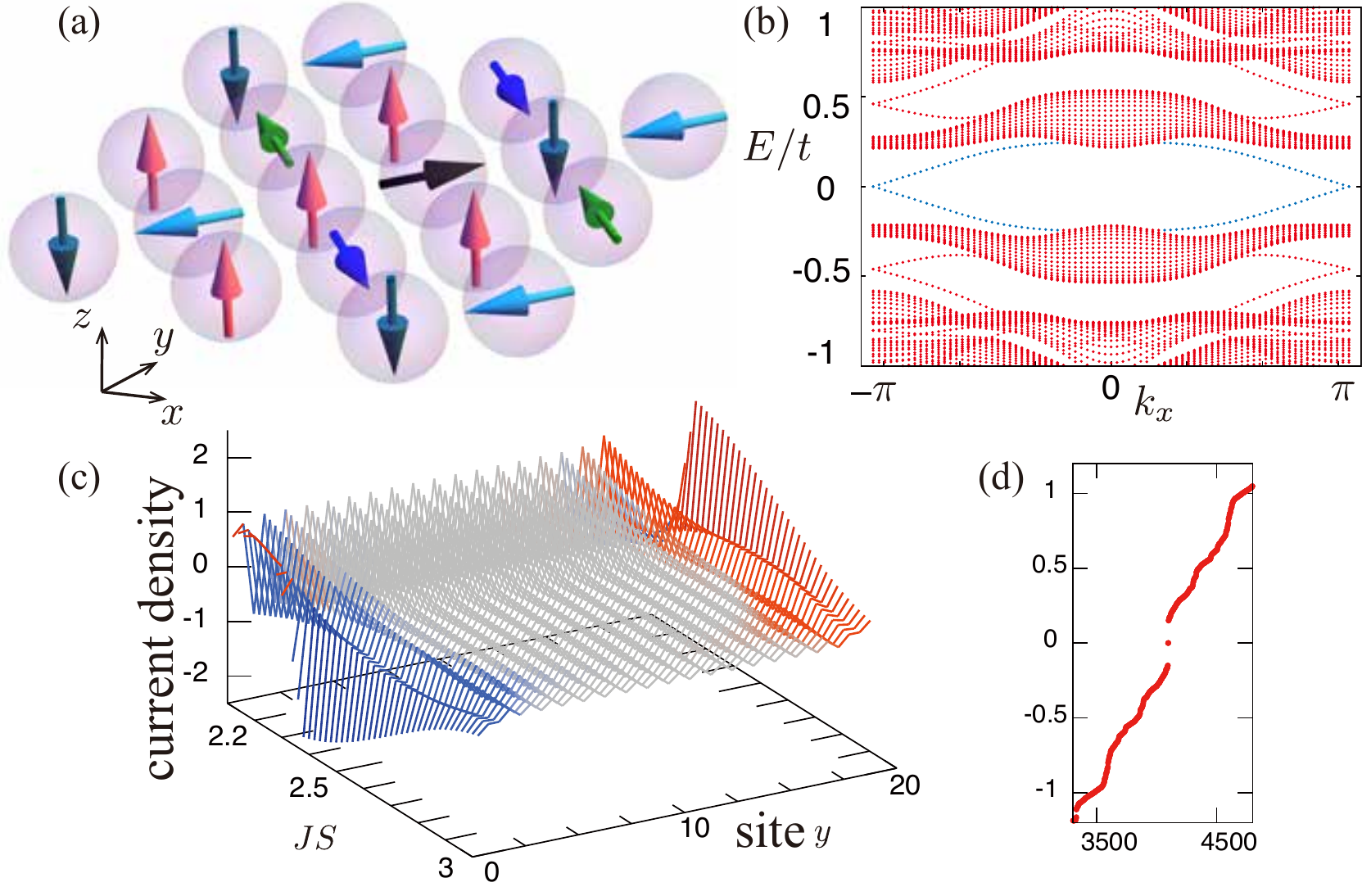}%
\caption{\label{fig:skx}
	(a) Configuration of skyrmion lattice producing an effective fully gapped superconducting phase 
		(See also Fig.~\ref{fig:fluc} in Supplemental Material~\ref{sec:fluc}). 
	(b) Energy spectrum of quasiparticles with the configuration in (a) calculated by tight binding model 
		(Eq.~(\ref{eq:tb})) with $\ t=1.0$ and $\ \Delta_0=0.7$ with open boundary condition along $y$-direction. 
		One can see linearly dispersing bands at $k_{x}=\pm \pi$ which are localized at the boundaries of the system.  
	(c) Current density calculated with the configuration given in (a) with 20 unit cells along $y$-direction. 
		One can see finite current density at the edges of the system, which is a crucial evidence for topologically 
		nontrivial phase. 
	(d)	Energy levels of quasiparticles obtained by a tight binding model calculation with vortices. The horizontal 
		axis shows the indices of energy eigenvalues. The system size is set $16\times16$ with periodic 
		boundary condition, and $t=\Delta_0=1.0$. The zero energy states are four-fold degenerate. 
	}
\end{figure}

2. \textit{Chiral $p$-wave superconductor.} 
Next we attempt to generate fully gapped superconducting states. For this purpose, the phase of pair potential 
along $x$- and $y$-direction should be different. The case where the phase difference is equal to $\pi/2$ is well 
known as chiral $p$-wave superconductor. We can always choose the phases of the pairing order parameter 
$\Delta_{ij}$ as real by appropriate gauge 
transformation once the moments lie in a plane. Then we need to consider non-coplanar spin configurations. 
Here we study the case of skyrmion crystal state (Fig.~\ref{fig:skx}~(a)) recently observed 
experimentally.~\cite{NatPhys.7.713, NatNano.8.152} 
In this case, moments 
obviously have a non-coplanar configuration. The parameters in the effective model calculated by Eqs.~(\ref{eq:t}) 
and~(\ref{eq:d}) are given in Supplemental Material~\ref{sec:td} (Fig.~\ref{fig:pskx}). We have confirmed the following 
properties of the system, and based on them we conclude that it has the same topological nature as chiral $p$-wave 
superconducting states while the transfer integrals and pair potentials are not uniform with this configuration. 
The characteristic properties of chiral $p$-wave superconductors are (i) the full gap nature, (ii) the existence of 
edge modes and currents, and (iii) the emergence of zero energy states at the cores of 
vortices.~\cite{JETPLett.70.609} 
First we have confirmed that the system has an energy gap in the whole Brillouin zone by the same calculation as 
Fig.~\ref{fig:helix}~(b), which indicates the complex value of the pair potential because the system is essentially 
spinless. Figure~\ref{fig:skx}~(b) shows the energy dispersion of quasiparticles with the open boundaries.  
One can see two linearly dispersing bands crossing at $k_{x}=\pm\pi$. 
which are localized at the edges of the system.~\cite{PhysRevB.64.054514} 
Here the Fermi surface in the normal state is hole-like, 
then chiral edge modes cross at $k_{x}=\pm\pi$ not $k_{x}=0$. 
The fact that these modes yield edge current is manifested by directly calculating current density
\begin{align}
j_{i} = \sum_{k_{x}\sigma} 2t \sin k_{x} \ \alpha^{\dagger}_{i\sigma}(k_{x}) \alpha_{i\sigma}(k_{x}).
\end{align}
The result is shown in Fig.~\ref{fig:skx}~(c). At $JS\sim2.4$ the clear signal appears indicating the topological 
quantum phase transition. We also confirmed that zero energy Majorana bound states appear at the cores of 
vortices~(Fig.~\ref{fig:skx}~(d)).
We conclude that the resulting superconducting state is in the same topological phase as the chiral $p$-wave 
superconductor based on these observations. 
As another example of the non-coplanar spin configuration, we study the double spiral structure in 
Supplemental Material~\ref{sec:dhelix}, and found the similar $p_{x}+ip_{y}$-wave like state. 
%

%===========================================================
% CONCLUSION
%===========================================================
\textit{Discussion and conclusions.---}
In this paper, we have proposed a new way of creating effective two-dimensional unconventional superconductivity by 
local moments on the conventional spin-singlet $s$-wave superconductor. The non-collinear configurations of moments 
are essential to induce $p$-wave pairing. 
There is a hierarchical structure in energy scale, i.e., the original $s$-wave energy gap and that of the induced 
$p$-wave superconductivity.  Andreev bound states by a topological origin appear within the latter energy gap. 
Even the chiral $p_x+ip_y$-wave like pairing is realized by the non-coplanar configuration of moments, which shows 
the chiral Majorana edge channel with linear dispersion and zero energy Majorana bound states at the vortex cores. 
Moreover it indicates that we can create various kinds of superconducting states by choosing appropriate 
configurations of moments. For example, by changing the distance between the moments, one can tune the magnitudes 
of effective transfer integrals and pair potentials. Then, the anisotropy $|t_x/t_y|$ can be controlled to obtain 
the rich topological phases discussed in Ref.~\onlinecite{PhysRevB.86.100504}.
Here we briefly discuss the effect of self-energy correction and spin fluctuation for legitimizing our approach 
and results. 
The self-energy correction due to the dynamical quantum fluctuation of spins can be estimated as $\mathrm{Re} 
\Sigma (\varepsilon) \sim \lambda \varepsilon$ and $\mathrm{Im} \Sigma (\varepsilon) \sim \lambda 
\frac{\varepsilon^2}{\varepsilon_{\mathrm{F}}}$ where $\lambda = J^2S/(I\varepsilon_{\mathrm{F}})$ is the 
dimensionless coupling constant with $I$ being the exchange coupling between spins in the magnet, and 
$\varepsilon_{\mathrm{F}}$ the Fermi energy of the superconductor. This correction is tiny at small electron 
energy $\varepsilon$, and does not change the mini-gap structure.
Also we have confirmed numerically the robustness of the induced gap structure against the small (static) spin 
fluctuation as shown in Fig.~\ref{fig:fluc} in Supplemental Material~\ref{sec:fluc}. 
Though our model will simulate the interface of bulk superconductors and magnetic materials, we end with an account 
of another experimental realization of these proposals. To create intended patterns of magnetic moments, we can use 
atomic manipulation techniques using scanning tunneling microscopy.~\cite{Science.262.218, PhysRevLett.92.056803} 
The spin structure in organized magnetic impurities is also observed,~\cite{Science.312.1021, Science.335.196} 
although it is antiferromagnetic and cannot be utilized for our proposal. 

%===========================================================
% ACKNOWLEDGEMENT
%===========================================================
\begin{acknowledgments}
\textit{Acknowledgment.---}
S.~N. was supported by Grant-in-Aid for JSPS Fellows. This work was supported by Grant-in-Aid for Scientific 
Research (S) (Grant No.~24224009); the Funding Program for World-Leading Innovative RD on Science and Technology 
(FIRST Program); Strategic International Cooperative Program (Joint Research Type) from Japan Science and 
Technology Agency; Innovative Areas ``Topological Quantum Phenomena'' (Grant No. 22103005) from the Ministry 
of Education, Culture, Sports, Science, and Technology of Japan.
\end{acknowledgments}

%===========================================================
% REFERENCE
%===========================================================
%%%%%%  .bbl from here
%merlin.mbs apsrev4-1.bst 2010-07-25 4.21a (PWD, AO, DPC) hacked
%Control: key (0)
%Control: author (72) initials jnrlst
%Control: editor formatted (1) identically to author
%Control: production of article title (-1) disabled
%Control: page (0) single
%Control: year (1) truncated
%Control: production of eprint (0) enabled
%
%%%%%%  .bbl up to here

\balancecolsandclearpage
\appendix
\renewcommand\theequation{S\arabic{equation}}
\renewcommand\thefigure{S\arabic{figure}}
\renewcommand\thesmsection{S\arabic{smsection}}
\renewcommand\thesmfig{S\arabic{smfig}}
\setcounter{equation}{0}
\setcounter{figure}{0}
\setcounter{section}{0}
\onecolumngrid

\begin{center}
\textbf{
{\normalsize Supplemental Material for \\
``Two-dimensional $p$-wave superconducting states with magnetic moments\\
 on a conventional $s$-wave superconductor''}}
 \end{center}

\section{$\bar{t}_{ij}$ and $\bar{\Delta}_{ij}$}
\label{sec:td}
\begin{figure}[ht]
\refstepcounter{smfig}\label{fig:phelix}
\centering
\begin{minipage}{.7\hsize}
\includegraphics[width=.75\hsize]{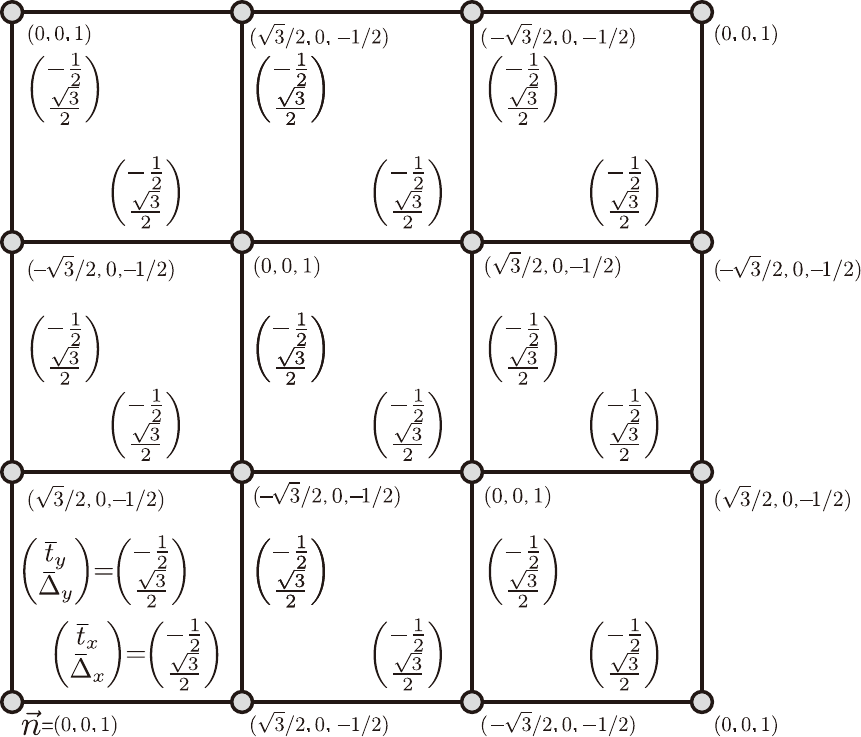}%
\caption{
	Spin configuration and parameters in the effective model for coplanar spin helix configuration (corresponding 
	to $p_{x}+p_{y}$-wave pairing states). $\bar{t}_{x}$ and 
	$\bar{t}_{y}$ are measured in the unit of $\bar{t}_0$ while $\bar{\Delta}_{x}$ and $\bar{\Delta}_{y}$ in the 
	unit of $\bar{\Delta}_0$ defined below Eq.~(\ref{eq:d}) in the main text. $\vec{n}=\vec{S}/S$ shows the 
	direction of the moment at each site. 
}
\end{minipage}
\end{figure}

\begin{figure}[hb]
\refstepcounter{smfig}\label{fig:pskx}
\centering
\begin{minipage}{.7\hsize}
\includegraphics[width=.9\hsize]{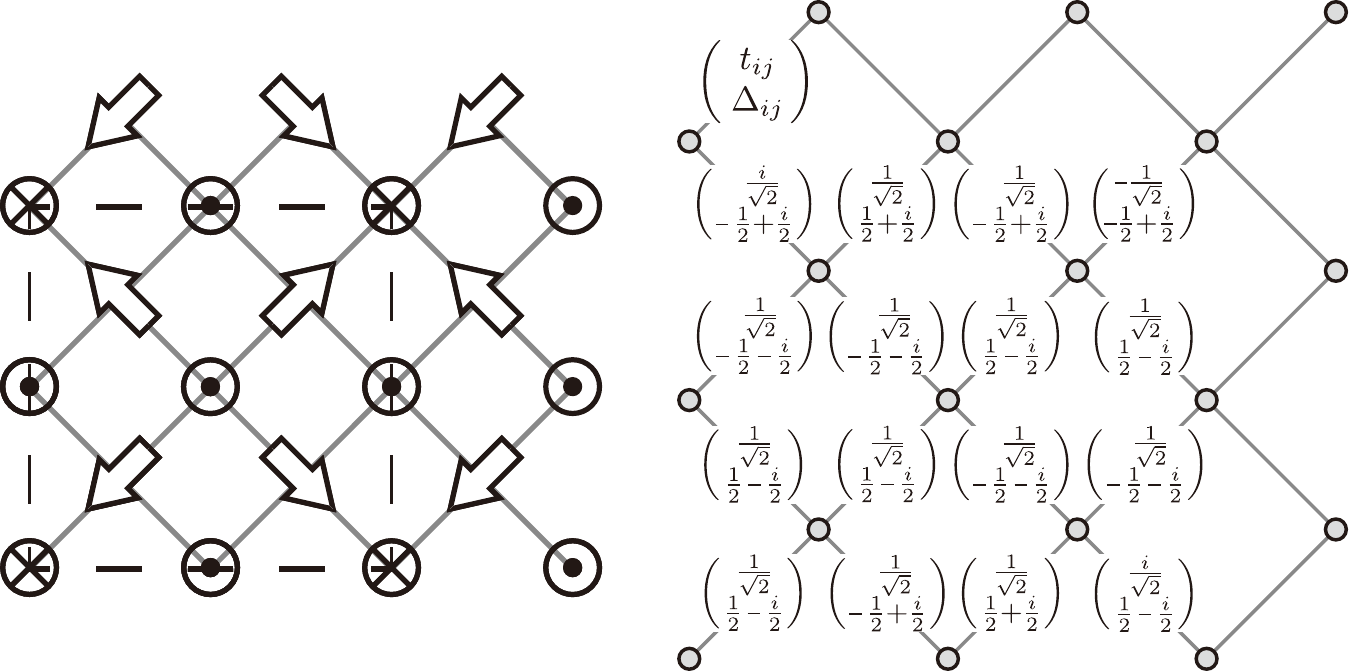}%
\caption{
	Spin configuration and parameters in the effective model for skyrmion lattice (corresponding to 
	$p_{x}+ip_{y}$-wave like pairing states). Here the spin configuration is given in a graphical way, which is 
	identical to Fig.~\ref{fig:skx}~(a). The arrows, circles with x (into the paper) and filled circle (out of the 
	paper) inside indicate the directions of the moments. Broken lines on right side show the magnetic unit cell.
	We take the same units as in Fig.~\ref{fig:phelix}.
}
\end{minipage}
\end{figure}

\begin{figure}[h]
\refstepcounter{smfig}\label{fig:pdhelix}
\centering
\begin{minipage}{.7\hsize}
\includegraphics[width=.75\hsize]{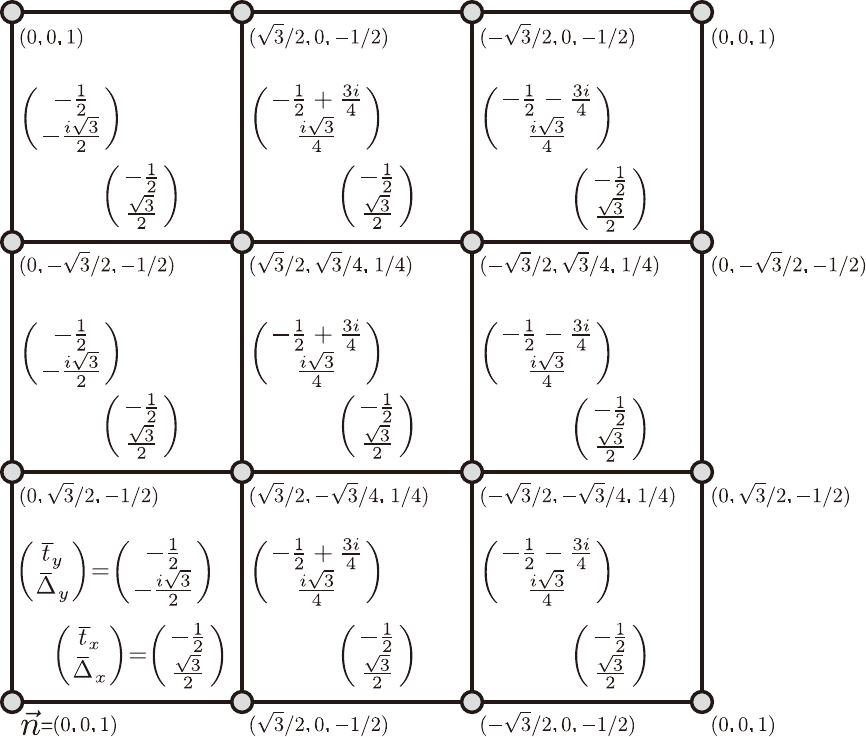}%
\caption{
	Spin configuration and parameters in the effective model for non-coplanar spin helix configuration (corresponding 
	to $p_{x}+ip_{y}$-wave like pairing states). We take the same units as in Fig.~\ref{fig:phelix}.
}
\end{minipage}
\end{figure}

We gave three particular spin configurations shown in Figs.~\ref{fig:helix}~(a), \ref{fig:skx}~(a), 
and~\ref{fig:dhelix}~(a).
Here we show the explicit values of the effective parameters calculated from Eqs.~(\ref{eq:t})~and~(\ref{eq:d}).
Figures~\ref{fig:phelix}, \ref{fig:pskx}, and \ref{fig:pdhelix} (for coplanar spin helix, skyrmion crystal, and non-coplanar 
spin helix, respectively) show following information; (i) direction $\vec{n}$ of the moment at each site in a unit 
cell, (ii) effective transfer integrals $\bar{t}$ and in the unit of $\bar{t}_0$, and 
(iii) effective pair potentials $\bar{\Delta}$ in the unit of $\bar{\Delta}_0$. 
The directions of moments are given by $\vec{n}=(n_{x},n_{y},n_{z})$ in Figs.~\ref{fig:phelix} and~\ref{fig:pdhelix}, and 
in a graphical way in Fig.~\ref{fig:pskx}.
The unit cell is $3\times3$ block for both spin helix configurations, and skyrmion crystal contains 8 atomic 
sites in a unit cell. 
$\bar{t}_0$ and $\bar{\Delta}_0$ are defined just below Eq.~(\ref{eq:d}) in the main text.

\section{Non-coplanar helical spin configuration}
\label{sec:dhelix}
\begin{figure}[h]
\refstepcounter{smfig}\label{fig:dhelix}
\centering
\begin{minipage}{.7\hsize}
\includegraphics[width=.9\hsize]{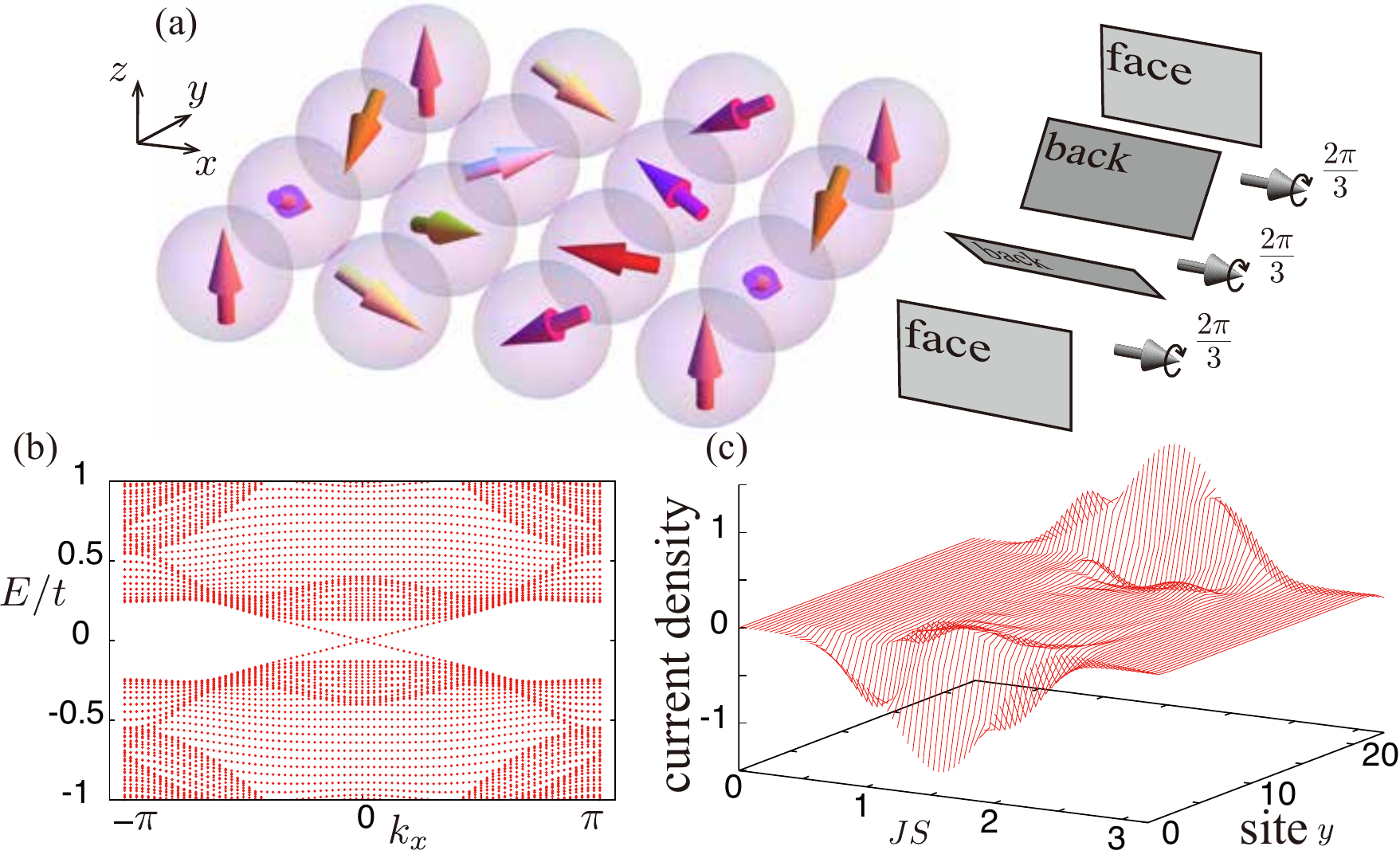}%
\caption{
	(a) Spin configuration producing effective chiral $p$-wave pairing. They have a non-coplanar structure. 
		The moments along $x$-direction rotate by $2\pi/3$ within the planes in the spin space schematically shown 
		on the right side. 
		This plane is rotated along the $y$-direction by $2\pi/3$ around the $s_x$-axis indicated by the arrows.
		Therefore $3\times3$ sites constitute a unit cell. 
	(b) Energy spectrum of quasiparticles with the periodic (open) boundary conditions along $x$-($y$-)direction. 
	(c) Current density calculated with the configuration given in (a) with the width $N_y=24$. 
		The other parameters are the same as in (b). There are finite chiral current density at the edges. 
}
\end{minipage}
\end{figure}

In the main text, we discuss the realization of the $p_{x}+ip_{y}$-wave like state by means of the skyrmion 
structure. Here in Supplemental Material we consider another non-coplanar configuration of magnetic moments. 
It is shown in Fig.~\ref{fig:dhelix}~(a). This can be seen as follows; we have sequential spin helix stripes along 
$x$-direction, but the planes on which spins rotate are tilted along $y$-direction. Describing in more detail, 
we start from the moment at the origin pointing along $+s_{z}$-direction, we rotate it by $2\pi/3$ within 
$s_{x}s_{z}$-plane along $x$-direction. Now we have a row of moments extended in $x$-direction. Then we rotate 
the plane where the moments lie by $2\pi/3$ around $s_{x}$-axis along $y$-direction. After three successive 
rotations, the plane comes back to the original one, i.e., $s_{x}s_{z}$-plane as shown in Fig.~\ref{fig:dhelix}~(a) 
right side. It is expected that the former rotation gives real $\bar{\Delta}_{ij}$'s in Eq.~(\ref{eq:d}) while 
the latter imaginary ones. 
The resulting $\bar{t}_{ij}$ and $\bar{\Delta}_{ij}$ are explicitly given in Fig.~\ref{fig:pdhelix}.
In fact, $\bar{t}_{i+x,i}=-\frac{1}{2}\bar{t}_0$ and $\bar{\Delta}_{i+x,i}=\frac{\sqrt{3}}{2}\bar{\Delta}_0$, 
which are consistent with the invariance of them under common rotation of two spins. 
On the other hand, along $y$-direction the moments have finite $s_{y}$ 
component, and hence complex $\bar{\Delta}_{i+y,i}$ results. Actually, they have pure imaginary values while 
$\bar{t}_{i+y,i}$ are not necessarily real. 
Note that any coplanar configurations of moments are transformed onto $s_{x}s_{z}$-plane by a common rotation, and 
all $\bar{t}_{ij}$ and $\bar{\Delta}_{ij}$ are essentially real.
To obtain complex $\bar{\Delta}_{ij}$'s, the non-coplanar configurations of moments are indispensable. 
While the transfer integrals and pair potentials in the effective model are not uniform with this 
configuration (see Fig.~\ref{fig:pdhelix}), it has the same topological nature as chiral $p$-wave superconductors 
as we will discuss below in the completely same ways as we do in the main text for the skyrmion structure. 

\begin{figure}[h]
\refstepcounter{smfig}\label{fig:vortex}
\centering
\begin{minipage}{.7\hsize}
\includegraphics[width=.8\hsize]{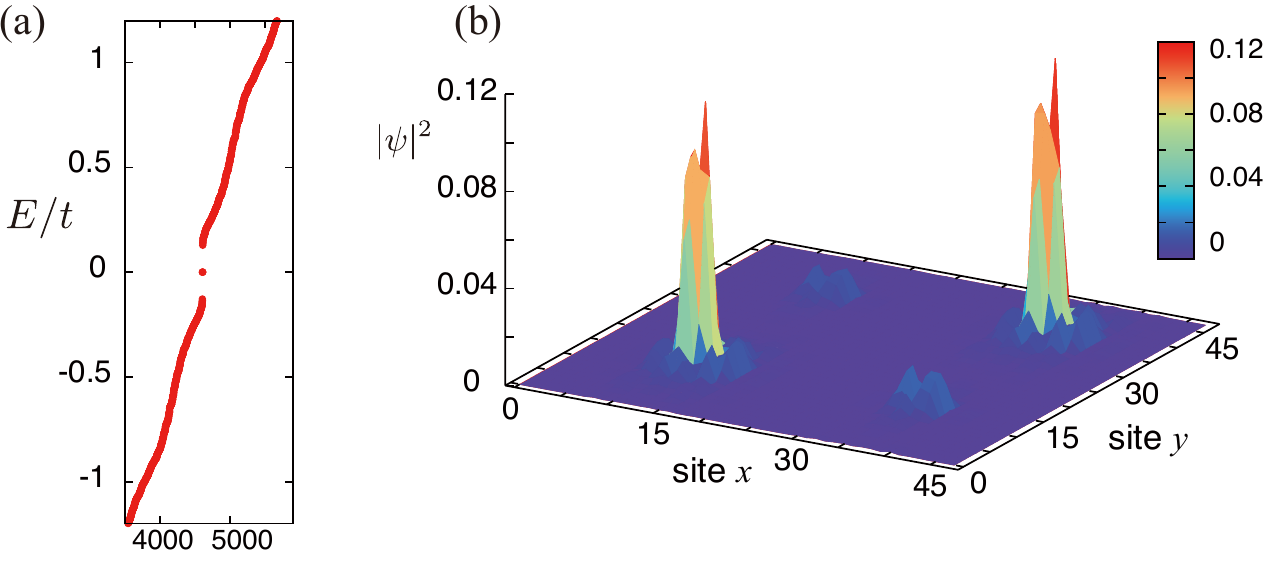}%
\caption{
	(a)	Energy levels of quasiparticles obtained by a tight binding model calculation with vortices. The horizontal 
		axis shows the indices of energy eigenvalues. The system size is set $48\times48$ with periodic 
		boundary condition, and $t=\Delta_0=1.0$. The zero energy states are four-fold degenerate. 
	(b) Distribution of a wave function of one of the zero energy states. It has peaks at the positions of vortex 
		cores. 
}
\end{minipage}
\end{figure}

First, the system described by the Hamiltonian in Eq.~(\ref{eq:tb}) has an energy gap in the whole Brillouin zone, 
which suggests complex value of pair potential. 
Figure~\ref{fig:dhelix}~(b) shows the energy dispersion of quasiparticles with the periodic (open) boundary condition 
along $x$- ($y$-)direction. The magnitude of $JS$ is chosen to be $2t$ so that the energy of the bound state 
around a single moment is about zero energy (see Fig.~\ref{fig:idea}). There are two one-dimensional bands as chiral 
Majorana edge channels localized at the open boundaries in the energy gap. 
To confirm the emergence of edge modes, we also calculate current density along $x$-direction
with site indices $i$'s along $y$-direction. The result is given in Fig.~\ref{fig:dhelix} (c), which clearly shows the 
presence of the edge current. 
It also shows that the signature of the topological quantum phase transition at $JS=(JS)_{c}\sim2.7t$. Namely, 
the system enters into a trivial phase for $JS>(JS)_{c}$ where the edge modes and current vanish. 
We have also confirmed that zero energy Majorana bound state appears at the cores of a vortices, which is consistent 
with chiral $p$-wave superconductors. 
We insert vortices into the pair potential of the original $s$-wave 
superconductor. To avoid the effect of the boundary which may cause in-gap states, we impose the periodic boundary 
condition for both $x$- and $y$-directions. For the consistency of the phase, we introduce two vortices with 
winding number 1 and other two anti-vortices with winding number -1. 
The four-fold zero energy states appear in the induced 
gap corresponding to the four vortices in total, and the probability of one of their wave functions in the 
real space shows localized distribution around the vortices as shown in Fig.~\ref{fig:vortex}. 
These results are the same as those in the case of the genuine chiral $p$-wave superconductor. 

In addition we study the effect of a defect. Here we use ``defect'' in the meaning of taking away one 
magnetic moment from the suite of aligned magnetic moments. In the tight binding picture, there is a lattice site 
for the original $s$-wave superconductor at the defect point. 
In the conventional $s$-wave superconductor, the non-magnetic defect does not produce any in-gap state. 
In sharp contrast, we find that new states appear inside the lower energy gap. Note that the energy is not 
necessarily zero. This property is characteristic to unconventional superconductors.~\cite{JPSJ.68.3054} 

We employ the rotation angle $2\pi/3$ in the coplanar and non-coplanar spin helix structure for the sake of 
numerical calculation. However, the properties discussed above obviously hold when we consider other angles as long 
as it is not equal to $0$ or $\pi$.

\section{Robustness against fluctuation}
\label{sec:fluc}
\begin{figure}[hb]
\refstepcounter{smfig}\label{fig:fluc}
\centering
\begin{minipage}{.7\hsize}
\includegraphics[width=.7\hsize]{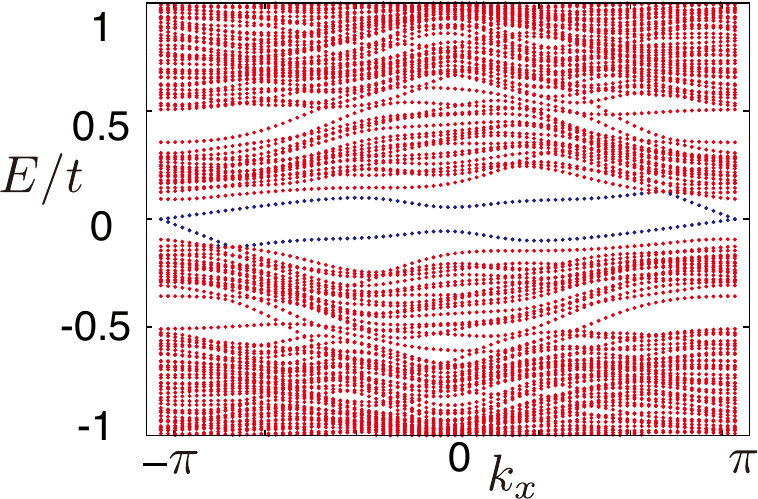}%
\caption{
	Energy spectrum with random modulation of directions of magnetic moments. The lines near energy 0 (blue ones) 
	are edge modes. They remain crossing at $k_{x}=\pm\pi$ and separated from the bulk states. 
}
\end{minipage}
\end{figure}

As stated in the main text, we assume the configuration of magnetic moments are strictly fixed all through the 
calculations. Here we examine the influence of spin fluctuation. It is shown that the results obtained above 
are qualitatively unaffected. 
To take into account the effect of spin fluctuation, we mimic it by modulating directions of the moments with 
random magnitude about $7\%$ from the static configuration (Fig.~\ref{fig:skx}~(a)). In Fig.~\ref{fig:fluc} the edge 
modes still connect occupied and unoccupied states even though they are slightly modified by the fluctuation. 

\begin{thebibliography}{50}%
\makeatletter
\providecommand \@ifxundefined [1]{%
 \@ifx{#1\undefined}
}%
\providecommand \@ifnum [1]{%
 \ifnum #1\expandafter \@firstoftwo
 \else \expandafter \@secondoftwo
 \fi
}%
\providecommand \@ifx [1]{%
 \ifx #1\expandafter \@firstoftwo
 \else \expandafter \@secondoftwo
 \fi
}%
\providecommand \natexlab [1]{#1}%
\providecommand \enquote  [1]{``#1''}%
\providecommand \bibnamefont  [1]{#1}%
\providecommand \bibfnamefont [1]{#1}%
\providecommand \citenamefont [1]{#1}%
\providecommand \href@noop [0]{\@secondoftwo}%
\providecommand \href [0]{\begingroup \@sanitize@url \@href}%
\providecommand \@href[1]{\@@startlink{#1}\@@href}%
\providecommand \@@href[1]{\endgroup#1\@@endlink}%
\providecommand \@sanitize@url [0]{\catcode `\\12\catcode `\$12\catcode
  `\&12\catcode `\#12\catcode `\^12\catcode `\_12\catcode `\%12\relax}%
\providecommand \@@startlink[1]{}%
\providecommand \@@endlink[0]{}%
\providecommand \url  [0]{\begingroup\@sanitize@url \@url }%
\providecommand \@url [1]{\endgroup\@href {#1}{\urlprefix }}%
\providecommand \urlprefix  [0]{URL }%
\providecommand \Eprint [0]{\href }%
\providecommand \doibase [0]{http://dx.doi.org/}%
\providecommand \selectlanguage [0]{\@gobble}%
\providecommand \bibinfo  [0]{\@secondoftwo}%
\providecommand \bibfield  [0]{\@secondoftwo}%
\providecommand \translation [1]{[#1]}%
\providecommand \BibitemOpen [0]{}%
\providecommand \bibitemStop [0]{}%
\providecommand \bibitemNoStop [0]{.\EOS\space}%
\providecommand \EOS [0]{\spacefactor3000\relax}%
\providecommand \BibitemShut  [1]{\csname bibitem#1\endcsname}%
\let\auto@bib@innerbib\@empty
%</preamble>
\bibitem [{\citenamefont {Sigrist}\ and\ \citenamefont
  {Ueda}(1991)}]{RevModPhys.63.239}%
  \BibitemOpen
  \bibfield  {author} {\bibinfo {author} {\bibfnamefont {M.}~\bibnamefont
  {Sigrist}}\ and\ \bibinfo {author} {\bibfnamefont {K.}~\bibnamefont {Ueda}},\
  }\href {\doibase 10.1103/RevModPhys.63.239} {\bibfield  {journal} {\bibinfo
  {journal} {Rev. Mod. Phys.}\ }\textbf {\bibinfo {volume} {63}},\ \bibinfo
  {pages} {239} (\bibinfo {year} {1991})}\BibitemShut {NoStop}%
\bibitem [{\citenamefont {Steglich}\ \emph {et~al.}(1979)\citenamefont
  {Steglich}, \citenamefont {Aarts}, \citenamefont {Bredl}, \citenamefont
  {Lieke}, \citenamefont {Meschede}, \citenamefont {Franz},\ and\ \citenamefont
  {Sch\"afer}}]{PhysRevLett.43.1892}%
  \BibitemOpen
  \bibfield  {author} {\bibinfo {author} {\bibfnamefont {F.}~\bibnamefont
  {Steglich}}, \bibinfo {author} {\bibfnamefont {J.}~\bibnamefont {Aarts}},
  \bibinfo {author} {\bibfnamefont {C.~D.}\ \bibnamefont {Bredl}}, \bibinfo
  {author} {\bibfnamefont {W.}~\bibnamefont {Lieke}}, \bibinfo {author}
  {\bibfnamefont {D.}~\bibnamefont {Meschede}}, \bibinfo {author}
  {\bibfnamefont {W.}~\bibnamefont {Franz}}, \ and\ \bibinfo {author}
  {\bibfnamefont {H.}~\bibnamefont {Sch\"afer}},\ }\href {\doibase
  10.1103/PhysRevLett.43.1892} {\bibfield  {journal} {\bibinfo  {journal}
  {Phys. Rev. Lett.}\ }\textbf {\bibinfo {volume} {43}},\ \bibinfo {pages}
  {1892} (\bibinfo {year} {1979})}\BibitemShut {NoStop}%
\bibitem [{\citenamefont {Tanaka}\ \emph {et~al.}(2012)\citenamefont {Tanaka},
  \citenamefont {Sato},\ and\ \citenamefont {Nagaosa}}]{JPSJ.81.011013}%
  \BibitemOpen
  \bibfield  {author} {\bibinfo {author} {\bibfnamefont {Y.}~\bibnamefont
  {Tanaka}}, \bibinfo {author} {\bibfnamefont {M.}~\bibnamefont {Sato}}, \ and\
  \bibinfo {author} {\bibfnamefont {N.}~\bibnamefont {Nagaosa}},\ }\href
  {\doibase 10.1143/JPSJ.81.011013} {\bibfield  {journal} {\bibinfo  {journal}
  {J. Phys. Soc. Jpn.}\ }\textbf {\bibinfo {volume} {81}},\ \bibinfo {pages}
  {011013} (\bibinfo {year} {2012})}\BibitemShut {NoStop}%
\bibitem [{\citenamefont {Read}\ and\ \citenamefont
  {Green}(2000)}]{PhysRevB.61.10267}%
  \BibitemOpen
  \bibfield  {author} {\bibinfo {author} {\bibfnamefont {N.}~\bibnamefont
  {Read}}\ and\ \bibinfo {author} {\bibfnamefont {D.}~\bibnamefont {Green}},\
  }\href {\doibase 10.1103/PhysRevB.61.10267} {\bibfield  {journal} {\bibinfo
  {journal} {Phys. Rev. B}\ }\textbf {\bibinfo {volume} {61}},\ \bibinfo
  {pages} {10267} (\bibinfo {year} {2000})}\BibitemShut {NoStop}%
\bibitem [{\citenamefont {Ivanov}(2001)}]{PhysRevLett.86.268}%
  \BibitemOpen
  \bibfield  {author} {\bibinfo {author} {\bibfnamefont {D.~A.}\ \bibnamefont
  {Ivanov}},\ }\href {\doibase 10.1103/PhysRevLett.86.268} {\bibfield
  {journal} {\bibinfo  {journal} {Phys. Rev. Lett.}\ }\textbf {\bibinfo
  {volume} {86}},\ \bibinfo {pages} {268} (\bibinfo {year} {2001})}\BibitemShut
  {NoStop}%
\bibitem [{\citenamefont {Moore}\ and\ \citenamefont
  {Read}(1991)}]{NuclPhysB.360.362}%
  \BibitemOpen
  \bibfield  {author} {\bibinfo {author} {\bibfnamefont {G.}~\bibnamefont
  {Moore}}\ and\ \bibinfo {author} {\bibfnamefont {N.}~\bibnamefont {Read}},\
  }\href {\doibase DOI: 10.1016/0550-3213(91)90407-O} {\bibfield  {journal}
  {\bibinfo  {journal} {Nucl. Phys. B}\ }\textbf {\bibinfo {volume} {360}},\
  \bibinfo {pages} {362 } (\bibinfo {year} {1991})}\BibitemShut {NoStop}%
\bibitem [{\citenamefont {Nayak}\ \emph {et~al.}(2008)\citenamefont {Nayak},
  \citenamefont {Simon}, \citenamefont {Stern}, \citenamefont {Freedman},\ and\
  \citenamefont {Das~Sarma}}]{RevModPhys.80.1083}%
  \BibitemOpen
  \bibfield  {author} {\bibinfo {author} {\bibfnamefont {C.}~\bibnamefont
  {Nayak}}, \bibinfo {author} {\bibfnamefont {S.~H.}\ \bibnamefont {Simon}},
  \bibinfo {author} {\bibfnamefont {A.}~\bibnamefont {Stern}}, \bibinfo
  {author} {\bibfnamefont {M.}~\bibnamefont {Freedman}}, \ and\ \bibinfo
  {author} {\bibfnamefont {S.}~\bibnamefont {Das~Sarma}},\ }\href {\doibase
  10.1103/RevModPhys.80.1083} {\bibfield  {journal} {\bibinfo  {journal} {Rev.
  Mod. Phys.}\ }\textbf {\bibinfo {volume} {80}},\ \bibinfo {pages} {1083}
  (\bibinfo {year} {2008})}\BibitemShut {NoStop}%
\bibitem [{\citenamefont {Qi}\ and\ \citenamefont
  {Zhang}(2011)}]{RevModPhys.83.1057}%
  \BibitemOpen
  \bibfield  {author} {\bibinfo {author} {\bibfnamefont {X.-L.}\ \bibnamefont
  {Qi}}\ and\ \bibinfo {author} {\bibfnamefont {S.-C.}\ \bibnamefont {Zhang}},\
  }\href {\doibase 10.1103/RevModPhys.83.1057} {\bibfield  {journal} {\bibinfo
  {journal} {Rev. Mod. Phys.}\ }\textbf {\bibinfo {volume} {83}},\ \bibinfo
  {pages} {1057} (\bibinfo {year} {2011})}\BibitemShut {NoStop}%
\bibitem [{\citenamefont {Alicea}(2012)}]{RepProgPhys.75.076501}%
  \BibitemOpen
  \bibfield  {author} {\bibinfo {author} {\bibfnamefont {J.}~\bibnamefont
  {Alicea}},\ }\href {http://stacks.iop.org/0034-4885/75/i=7/a=076501}
  {\bibfield  {journal} {\bibinfo  {journal} {Rep. Prog. Phys.}\ }\textbf
  {\bibinfo {volume} {75}},\ \bibinfo {pages} {076501} (\bibinfo {year}
  {2012})}\BibitemShut {NoStop}%
\bibitem [{\citenamefont {Sau}\ \emph {et~al.}(2010)\citenamefont {Sau},
  \citenamefont {Lutchyn}, \citenamefont {Tewari},\ and\ \citenamefont
  {Das~Sarma}}]{PhysRevLett.104.040502}%
  \BibitemOpen
  \bibfield  {author} {\bibinfo {author} {\bibfnamefont {J.~D.}\ \bibnamefont
  {Sau}}, \bibinfo {author} {\bibfnamefont {R.~M.}\ \bibnamefont {Lutchyn}},
  \bibinfo {author} {\bibfnamefont {S.}~\bibnamefont {Tewari}}, \ and\ \bibinfo
  {author} {\bibfnamefont {S.}~\bibnamefont {Das~Sarma}},\ }\href {\doibase
  10.1103/PhysRevLett.104.040502} {\bibfield  {journal} {\bibinfo  {journal}
  {Phys. Rev. Lett.}\ }\textbf {\bibinfo {volume} {104}},\ \bibinfo {pages}
  {040502} (\bibinfo {year} {2010})}\BibitemShut {NoStop}%
\bibitem [{\citenamefont {Alicea}(2010)}]{PhysRevB.81.125318}%
  \BibitemOpen
  \bibfield  {author} {\bibinfo {author} {\bibfnamefont {J.}~\bibnamefont
  {Alicea}},\ }\href {\doibase 10.1103/PhysRevB.81.125318} {\bibfield
  {journal} {\bibinfo  {journal} {Phys. Rev. B}\ }\textbf {\bibinfo {volume}
  {81}},\ \bibinfo {pages} {125318} (\bibinfo {year} {2010})}\BibitemShut
  {NoStop}%
\bibitem [{\citenamefont {Lutchyn}\ \emph {et~al.}(2010)\citenamefont
  {Lutchyn}, \citenamefont {Sau},\ and\ \citenamefont
  {Das~Sarma}}]{PhysRevLett.105.077001}%
  \BibitemOpen
  \bibfield  {author} {\bibinfo {author} {\bibfnamefont {R.~M.}\ \bibnamefont
  {Lutchyn}}, \bibinfo {author} {\bibfnamefont {J.~D.}\ \bibnamefont {Sau}}, \
  and\ \bibinfo {author} {\bibfnamefont {S.}~\bibnamefont {Das~Sarma}},\ }\href
  {\doibase 10.1103/PhysRevLett.105.077001} {\bibfield  {journal} {\bibinfo
  {journal} {Phys. Rev. Lett.}\ }\textbf {\bibinfo {volume} {105}},\ \bibinfo
  {pages} {077001} (\bibinfo {year} {2010})}\BibitemShut {NoStop}%
\bibitem [{\citenamefont {Oreg}\ \emph {et~al.}(2010)\citenamefont {Oreg},
  \citenamefont {Refael},\ and\ \citenamefont {von
  Oppen}}]{PhysRevLett.105.177002}%
  \BibitemOpen
  \bibfield  {author} {\bibinfo {author} {\bibfnamefont {Y.}~\bibnamefont
  {Oreg}}, \bibinfo {author} {\bibfnamefont {G.}~\bibnamefont {Refael}}, \ and\
  \bibinfo {author} {\bibfnamefont {F.}~\bibnamefont {von Oppen}},\ }\href
  {\doibase 10.1103/PhysRevLett.105.177002} {\bibfield  {journal} {\bibinfo
  {journal} {Phys. Rev. Lett.}\ }\textbf {\bibinfo {volume} {105}},\ \bibinfo
  {pages} {177002} (\bibinfo {year} {2010})}\BibitemShut {NoStop}%
\bibitem [{\citenamefont {Stanescu}\ and\ \citenamefont
  {Tewari}(2013)}]{JPhysCondMatt.25.233201}%
  \BibitemOpen
  \bibfield  {author} {\bibinfo {author} {\bibfnamefont {T.~D.}\ \bibnamefont
  {Stanescu}}\ and\ \bibinfo {author} {\bibfnamefont {S.}~\bibnamefont
  {Tewari}},\ }\href {http://stacks.iop.org/0953-8984/25/i=23/a=233201}
  {\bibfield  {journal} {\bibinfo  {journal} {J. Phys.: Cond. Matt.}\ }\textbf
  {\bibinfo {volume} {25}},\ \bibinfo {pages} {233201} (\bibinfo {year}
  {2013})}\BibitemShut {NoStop}%
\bibitem [{\citenamefont {Mourik}\ \emph {et~al.}(2012)\citenamefont {Mourik},
  \citenamefont {Zuo}, \citenamefont {Frolov}, \citenamefont {Plissard},
  \citenamefont {Bakkers},\ and\ \citenamefont
  {Kouwenhoven}}]{Science.336.1003}%
  \BibitemOpen
  \bibfield  {author} {\bibinfo {author} {\bibfnamefont {V.}~\bibnamefont
  {Mourik}}, \bibinfo {author} {\bibfnamefont {K.}~\bibnamefont {Zuo}},
  \bibinfo {author} {\bibfnamefont {S.~M.}\ \bibnamefont {Frolov}}, \bibinfo
  {author} {\bibfnamefont {S.~R.}\ \bibnamefont {Plissard}}, \bibinfo {author}
  {\bibfnamefont {E.~P. A.~M.}\ \bibnamefont {Bakkers}}, \ and\ \bibinfo
  {author} {\bibfnamefont {L.~P.}\ \bibnamefont {Kouwenhoven}},\ }\href
  {\doibase 10.1126/science.1222360} {\bibfield  {journal} {\bibinfo  {journal}
  {Science}\ }\textbf {\bibinfo {volume} {336}},\ \bibinfo {pages} {1003}
  (\bibinfo {year} {2012})}\BibitemShut {NoStop}%
\bibitem [{\citenamefont {Deng}\ \emph {et~al.}(2012)\citenamefont {Deng},
  \citenamefont {Yu}, \citenamefont {Huang}, \citenamefont {Larsson},
  \citenamefont {Caroff},\ and\ \citenamefont {Xu}}]{NanoLett.12.6414}%
  \BibitemOpen
  \bibfield  {author} {\bibinfo {author} {\bibfnamefont {M.~T.}\ \bibnamefont
  {Deng}}, \bibinfo {author} {\bibfnamefont {C.~L.}\ \bibnamefont {Yu}},
  \bibinfo {author} {\bibfnamefont {G.~Y.}\ \bibnamefont {Huang}}, \bibinfo
  {author} {\bibfnamefont {M.}~\bibnamefont {Larsson}}, \bibinfo {author}
  {\bibfnamefont {P.}~\bibnamefont {Caroff}}, \ and\ \bibinfo {author}
  {\bibfnamefont {H.~Q.}\ \bibnamefont {Xu}},\ }\href {\doibase
  10.1021/nl303758w} {\bibfield  {journal} {\bibinfo  {journal} {Nano Letters}\
  }\textbf {\bibinfo {volume} {12}},\ \bibinfo {pages} {6414} (\bibinfo {year}
  {2012})}\BibitemShut {NoStop}%
\bibitem [{\citenamefont {Rokhinson}\ \emph {et~al.}(2012)\citenamefont
  {Rokhinson}, \citenamefont {Liu},\ and\ \citenamefont
  {Furdyna}}]{NatPhys.8.795}%
  \BibitemOpen
  \bibfield  {author} {\bibinfo {author} {\bibfnamefont {L.~P.}\ \bibnamefont
  {Rokhinson}}, \bibinfo {author} {\bibfnamefont {X.}~\bibnamefont {Liu}}, \
  and\ \bibinfo {author} {\bibfnamefont {J.~K.}\ \bibnamefont {Furdyna}},\
  }\href {\doibase 10.1038/nphys2429} {\bibfield  {journal} {\bibinfo
  {journal} {Nat. Phys.}\ }\textbf {\bibinfo {volume} {8}},\ \bibinfo {pages}
  {795} (\bibinfo {year} {2012})}\BibitemShut {NoStop}%
\bibitem [{\citenamefont {Das}\ \emph {et~al.}(2012)\citenamefont {Das},
  \citenamefont {Ronen}, \citenamefont {Most}, \citenamefont {Oreg},
  \citenamefont {Heiblum},\ and\ \citenamefont {Shtrikman}}]{NatPhys.8.887}%
  \BibitemOpen
  \bibfield  {author} {\bibinfo {author} {\bibfnamefont {A.}~\bibnamefont
  {Das}}, \bibinfo {author} {\bibfnamefont {Y.}~\bibnamefont {Ronen}}, \bibinfo
  {author} {\bibfnamefont {Y.}~\bibnamefont {Most}}, \bibinfo {author}
  {\bibfnamefont {Y.}~\bibnamefont {Oreg}}, \bibinfo {author} {\bibfnamefont
  {M.}~\bibnamefont {Heiblum}}, \ and\ \bibinfo {author} {\bibfnamefont
  {H.}~\bibnamefont {Shtrikman}},\ }\href {\doibase 10.1038/nphys2479}
  {\bibfield  {journal} {\bibinfo  {journal} {Nat. Phys.}\ }\textbf {\bibinfo
  {volume} {8}},\ \bibinfo {pages} {887} (\bibinfo {year} {2012})}\BibitemShut
  {NoStop}%
\bibitem [{\citenamefont {Fu}\ and\ \citenamefont
  {Kane}(2008)}]{PhysRevLett.100.096407}%
  \BibitemOpen
  \bibfield  {author} {\bibinfo {author} {\bibfnamefont {L.}~\bibnamefont
  {Fu}}\ and\ \bibinfo {author} {\bibfnamefont {C.~L.}\ \bibnamefont {Kane}},\
  }\href {\doibase 10.1103/PhysRevLett.100.096407} {\bibfield  {journal}
  {\bibinfo  {journal} {Phys. Rev. Lett.}\ }\textbf {\bibinfo {volume} {100}},\
  \bibinfo {pages} {096407} (\bibinfo {year} {2008})}\BibitemShut {NoStop}%
\bibitem [{\citenamefont {Linder}\ \emph {et~al.}(2010)\citenamefont {Linder},
  \citenamefont {Tanaka}, \citenamefont {Yokoyama}, \citenamefont {Sudb\o{}},\
  and\ \citenamefont {Nagaosa}}]{PhysRevLett.104.067001}%
  \BibitemOpen
  \bibfield  {author} {\bibinfo {author} {\bibfnamefont {J.}~\bibnamefont
  {Linder}}, \bibinfo {author} {\bibfnamefont {Y.}~\bibnamefont {Tanaka}},
  \bibinfo {author} {\bibfnamefont {T.}~\bibnamefont {Yokoyama}}, \bibinfo
  {author} {\bibfnamefont {A.}~\bibnamefont {Sudb\o{}}}, \ and\ \bibinfo
  {author} {\bibfnamefont {N.}~\bibnamefont {Nagaosa}},\ }\href {\doibase
  10.1103/PhysRevLett.104.067001} {\bibfield  {journal} {\bibinfo  {journal}
  {Phys. Rev. Lett.}\ }\textbf {\bibinfo {volume} {104}},\ \bibinfo {pages}
  {067001} (\bibinfo {year} {2010})}\BibitemShut {NoStop}%
\bibitem [{\citenamefont {Sato}\ \emph {et~al.}(2009)\citenamefont {Sato},
  \citenamefont {Takahashi},\ and\ \citenamefont
  {Fujimoto}}]{PhysRevLett.103.020401}%
  \BibitemOpen
  \bibfield  {author} {\bibinfo {author} {\bibfnamefont {M.}~\bibnamefont
  {Sato}}, \bibinfo {author} {\bibfnamefont {Y.}~\bibnamefont {Takahashi}}, \
  and\ \bibinfo {author} {\bibfnamefont {S.}~\bibnamefont {Fujimoto}},\ }\href
  {\doibase 10.1103/PhysRevLett.103.020401} {\bibfield  {journal} {\bibinfo
  {journal} {Phys. Rev. Lett.}\ }\textbf {\bibinfo {volume} {103}},\ \bibinfo
  {pages} {020401} (\bibinfo {year} {2009})}\BibitemShut {NoStop}%
\bibitem [{\citenamefont {Sau}\ and\ \citenamefont
  {Sarma}(2012)}]{NatCommun.3.964}%
  \BibitemOpen
  \bibfield  {author} {\bibinfo {author} {\bibfnamefont {J.~D.}\ \bibnamefont
  {Sau}}\ and\ \bibinfo {author} {\bibfnamefont {S.~D.}\ \bibnamefont
  {Sarma}},\ }\href {\doibase 10.1038/ncomms1966} {\bibfield  {journal}
  {\bibinfo  {journal} {Nat. Commun.}\ }\textbf {\bibinfo {volume} {3}},\
  \bibinfo {pages} {964} (\bibinfo {year} {2012})}\BibitemShut {NoStop}%
\bibitem [{\citenamefont {Braunecker}\ \emph {et~al.}(2010)\citenamefont
  {Braunecker}, \citenamefont {Japaridze}, \citenamefont {Klinovaja},\ and\
  \citenamefont {Loss}}]{PhysRevB.82.045127}%
  \BibitemOpen
  \bibfield  {author} {\bibinfo {author} {\bibfnamefont {B.}~\bibnamefont
  {Braunecker}}, \bibinfo {author} {\bibfnamefont {G.~I.}\ \bibnamefont
  {Japaridze}}, \bibinfo {author} {\bibfnamefont {J.}~\bibnamefont
  {Klinovaja}}, \ and\ \bibinfo {author} {\bibfnamefont {D.}~\bibnamefont
  {Loss}},\ }\href {\doibase 10.1103/PhysRevB.82.045127} {\bibfield  {journal}
  {\bibinfo  {journal} {Phys. Rev. B}\ }\textbf {\bibinfo {volume} {82}},\
  \bibinfo {pages} {045127} (\bibinfo {year} {2010})}\BibitemShut {NoStop}%
\bibitem [{\citenamefont {Choy}\ \emph {et~al.}(2011)\citenamefont {Choy},
  \citenamefont {Edge}, \citenamefont {Akhmerov},\ and\ \citenamefont
  {Beenakker}}]{PhysRevB.84.195442}%
  \BibitemOpen
  \bibfield  {author} {\bibinfo {author} {\bibfnamefont {T.-P.}\ \bibnamefont
  {Choy}}, \bibinfo {author} {\bibfnamefont {J.~M.}\ \bibnamefont {Edge}},
  \bibinfo {author} {\bibfnamefont {A.~R.}\ \bibnamefont {Akhmerov}}, \ and\
  \bibinfo {author} {\bibfnamefont {C.~W.~J.}\ \bibnamefont {Beenakker}},\
  }\href {\doibase 10.1103/PhysRevB.84.195442} {\bibfield  {journal} {\bibinfo
  {journal} {Phys. Rev. B}\ }\textbf {\bibinfo {volume} {84}},\ \bibinfo
  {pages} {195442} (\bibinfo {year} {2011})}\BibitemShut {NoStop}%
\bibitem [{\citenamefont {Martin}\ and\ \citenamefont
  {Morpurgo}(2012)}]{PhysRevB.85.144505}%
  \BibitemOpen
  \bibfield  {author} {\bibinfo {author} {\bibfnamefont {I.}~\bibnamefont
  {Martin}}\ and\ \bibinfo {author} {\bibfnamefont {A.~F.}\ \bibnamefont
  {Morpurgo}},\ }\href {\doibase 10.1103/PhysRevB.85.144505} {\bibfield
  {journal} {\bibinfo  {journal} {Phys. Rev. B}\ }\textbf {\bibinfo {volume}
  {85}},\ \bibinfo {pages} {144505} (\bibinfo {year} {2012})}\BibitemShut
  {NoStop}%
\bibitem [{\citenamefont {Nadj-Perge}\ \emph {et~al.}(2013)\citenamefont
  {Nadj-Perge}, \citenamefont {Drozdov}, \citenamefont {Bernevig},\ and\
  \citenamefont {Yazdani}}]{PhysRevB.88.020407}%
  \BibitemOpen
  \bibfield  {author} {\bibinfo {author} {\bibfnamefont {S.}~\bibnamefont
  {Nadj-Perge}}, \bibinfo {author} {\bibfnamefont {I.~K.}\ \bibnamefont
  {Drozdov}}, \bibinfo {author} {\bibfnamefont {B.~A.}\ \bibnamefont
  {Bernevig}}, \ and\ \bibinfo {author} {\bibfnamefont {A.}~\bibnamefont
  {Yazdani}},\ }\href {\doibase 10.1103/PhysRevB.88.020407} {\bibfield
  {journal} {\bibinfo  {journal} {Phys. Rev. B}\ }\textbf {\bibinfo {volume}
  {88}},\ \bibinfo {pages} {020407} (\bibinfo {year} {2013})}\BibitemShut
  {NoStop}%
\bibitem [{\citenamefont {Lu}\ and\ \citenamefont
  {Wang}(2013)}]{PhysRevLett.110.096403}%
  \BibitemOpen
  \bibfield  {author} {\bibinfo {author} {\bibfnamefont {Y.-M.}\ \bibnamefont
  {Lu}}\ and\ \bibinfo {author} {\bibfnamefont {Z.}~\bibnamefont {Wang}},\
  }\href {\doibase 10.1103/PhysRevLett.110.096403} {\bibfield  {journal}
  {\bibinfo  {journal} {Phys. Rev. Lett.}\ }\textbf {\bibinfo {volume} {110}},\
  \bibinfo {pages} {096403} (\bibinfo {year} {2013})}\BibitemShut {NoStop}%
\bibitem [{\citenamefont {Shiba}(1968)}]{ProgTheorPhys.40.435}%
  \BibitemOpen
  \bibfield  {author} {\bibinfo {author} {\bibfnamefont {H.}~\bibnamefont
  {Shiba}},\ }\href {\doibase 10.1143/PTP.40.435} {\bibfield  {journal}
  {\bibinfo  {journal} {Prog. Theor. Phys.}\ }\textbf {\bibinfo {volume}
  {40}},\ \bibinfo {pages} {435} (\bibinfo {year} {1968})}\BibitemShut
  {NoStop}%
\bibitem [{\citenamefont {Rusinov}(1969)}]{SovPhysJETP.29.1101}%
  \BibitemOpen
  \bibfield  {author} {\bibinfo {author} {\bibfnamefont {A.~I.}\ \bibnamefont
  {Rusinov}},\ }\href
  {http://www.jetp.ac.ru/cgi-bin/e/index/e/29/6/p1101?a=list} {\bibfield
  {journal} {\bibinfo  {journal} {Sov. Phys. JETP}\ }\textbf {\bibinfo {volume}
  {29}},\ \bibinfo {pages} {1101} (\bibinfo {year} {1969})}\BibitemShut
  {NoStop}%
\bibitem [{\citenamefont {Yu}(1985)}]{ActaPhysSin.21.75}%
  \BibitemOpen
  \bibfield  {author} {\bibinfo {author} {\bibfnamefont {L.}~\bibnamefont
  {Yu}},\ }\href {http://wulixb.iphy.ac.cn/EN/abstract/abstract851.shtml}
  {\bibfield  {journal} {\bibinfo  {journal} {Acta Phys. Sin.}\ }\textbf
  {\bibinfo {volume} {21}},\ \bibinfo {pages} {75} (\bibinfo {year}
  {1985})}\BibitemShut {NoStop}%
\bibitem [{\citenamefont {Yazdani}\ \emph {et~al.}(1997)\citenamefont
  {Yazdani}, \citenamefont {Jones}, \citenamefont {Lutz}, \citenamefont
  {Crommie},\ and\ \citenamefont {Eigler}}]{Science.275.1767}%
  \BibitemOpen
  \bibfield  {author} {\bibinfo {author} {\bibfnamefont {A.}~\bibnamefont
  {Yazdani}}, \bibinfo {author} {\bibfnamefont {B.~A.}\ \bibnamefont {Jones}},
  \bibinfo {author} {\bibfnamefont {C.~P.}\ \bibnamefont {Lutz}}, \bibinfo
  {author} {\bibfnamefont {M.~F.}\ \bibnamefont {Crommie}}, \ and\ \bibinfo
  {author} {\bibfnamefont {D.~M.}\ \bibnamefont {Eigler}},\ }\href {\doibase
  10.1126/science.275.5307.1767} {\bibfield  {journal} {\bibinfo  {journal}
  {Science}\ }\textbf {\bibinfo {volume} {275}},\ \bibinfo {pages} {1767}
  (\bibinfo {year} {1997})}\BibitemShut {NoStop}%
\bibitem [{\citenamefont {Buzdin}(2005)}]{RevModPhys.77.935}%
  \BibitemOpen
  \bibfield  {author} {\bibinfo {author} {\bibfnamefont {A.~I.}\ \bibnamefont
  {Buzdin}},\ }\href {\doibase 10.1103/RevModPhys.77.935} {\bibfield  {journal}
  {\bibinfo  {journal} {Rev. Mod. Phys.}\ }\textbf {\bibinfo {volume} {77}},\
  \bibinfo {pages} {935} (\bibinfo {year} {2005})}\BibitemShut {NoStop}%
\bibitem [{\citenamefont {Bergeret}\ \emph {et~al.}(2005)\citenamefont
  {Bergeret}, \citenamefont {Volkov},\ and\ \citenamefont
  {Efetov}}]{RevModPhys.77.1321}%
  \BibitemOpen
  \bibfield  {author} {\bibinfo {author} {\bibfnamefont {F.~S.}\ \bibnamefont
  {Bergeret}}, \bibinfo {author} {\bibfnamefont {A.~F.}\ \bibnamefont
  {Volkov}}, \ and\ \bibinfo {author} {\bibfnamefont {K.~B.}\ \bibnamefont
  {Efetov}},\ }\href {\doibase 10.1103/RevModPhys.77.1321} {\bibfield
  {journal} {\bibinfo  {journal} {Rev. Mod. Phys.}\ }\textbf {\bibinfo {volume}
  {77}},\ \bibinfo {pages} {1321} (\bibinfo {year} {2005})}\BibitemShut
  {NoStop}%
\bibitem [{\citenamefont {Linder}\ \emph {et~al.}(2009)\citenamefont {Linder},
  \citenamefont {Yokoyama},\ and\ \citenamefont
  {Sudb\o{}}}]{PhysRevB.79.054523}%
  \BibitemOpen
  \bibfield  {author} {\bibinfo {author} {\bibfnamefont {J.}~\bibnamefont
  {Linder}}, \bibinfo {author} {\bibfnamefont {T.}~\bibnamefont {Yokoyama}}, \
  and\ \bibinfo {author} {\bibfnamefont {A.}~\bibnamefont {Sudb\o{}}},\ }\href
  {\doibase 10.1103/PhysRevB.79.054523} {\bibfield  {journal} {\bibinfo
  {journal} {Phys. Rev. B}\ }\textbf {\bibinfo {volume} {79}},\ \bibinfo
  {pages} {054523} (\bibinfo {year} {2009})}\BibitemShut {NoStop}%
\bibitem [{\citenamefont {Fert}\ \emph {et~al.}(2013)\citenamefont {Fert},
  \citenamefont {Cros},\ and\ \citenamefont {Sampaio}}]{NatNano.8.152}%
  \BibitemOpen
  \bibfield  {author} {\bibinfo {author} {\bibfnamefont {A.}~\bibnamefont
  {Fert}}, \bibinfo {author} {\bibfnamefont {V.}~\bibnamefont {Cros}}, \ and\
  \bibinfo {author} {\bibfnamefont {J.}~\bibnamefont {Sampaio}},\ }\href
  {http://dx.doi.org/10.1038/nnano.2013.29} {\bibfield  {journal} {\bibinfo
  {journal} {Nat. Nano.}\ }\textbf {\bibinfo {volume} {8}},\ \bibinfo {pages}
  {152} (\bibinfo {year} {2013})}\BibitemShut {NoStop}%
\bibitem [{\citenamefont {Heinze}\ \emph {et~al.}(2011)\citenamefont {Heinze},
  \citenamefont {von Bergmann}, \citenamefont {Menzel}, \citenamefont {Brede},
  \citenamefont {Kubetzka}, \citenamefont {Wiesendanger}, \citenamefont
  {Bihlmayer},\ and\ \citenamefont {Blugel}}]{NatPhys.7.713}%
  \BibitemOpen
  \bibfield  {author} {\bibinfo {author} {\bibfnamefont {S.}~\bibnamefont
  {Heinze}}, \bibinfo {author} {\bibfnamefont {K.}~\bibnamefont {von
  Bergmann}}, \bibinfo {author} {\bibfnamefont {M.}~\bibnamefont {Menzel}},
  \bibinfo {author} {\bibfnamefont {J.}~\bibnamefont {Brede}}, \bibinfo
  {author} {\bibfnamefont {A.}~\bibnamefont {Kubetzka}}, \bibinfo {author}
  {\bibfnamefont {R.}~\bibnamefont {Wiesendanger}}, \bibinfo {author}
  {\bibfnamefont {G.}~\bibnamefont {Bihlmayer}}, \ and\ \bibinfo {author}
  {\bibfnamefont {S.}~\bibnamefont {Blugel}},\ }\href
  {http://dx.doi.org/10.1038/nphys2045} {\bibfield  {journal} {\bibinfo
  {journal} {Nat Phys}\ }\textbf {\bibinfo {volume} {7}},\ \bibinfo {pages}
  {713} (\bibinfo {year} {2011})}\BibitemShut {NoStop}%
\bibitem [{\citenamefont {Balatsky}\ \emph {et~al.}(2006)\citenamefont
  {Balatsky}, \citenamefont {Vekhter},\ and\ \citenamefont
  {Zhu}}]{RevModPhys.78.373}%
  \BibitemOpen
  \bibfield  {author} {\bibinfo {author} {\bibfnamefont {A.~V.}\ \bibnamefont
  {Balatsky}}, \bibinfo {author} {\bibfnamefont {I.}~\bibnamefont {Vekhter}}, \
  and\ \bibinfo {author} {\bibfnamefont {J.-X.}\ \bibnamefont {Zhu}},\ }\href
  {\doibase 10.1103/RevModPhys.78.373} {\bibfield  {journal} {\bibinfo
  {journal} {Rev. Mod. Phys.}\ }\textbf {\bibinfo {volume} {78}},\ \bibinfo
  {pages} {373} (\bibinfo {year} {2006})}\BibitemShut {NoStop}%
\bibitem [{\citenamefont {Hara}\ and\ \citenamefont
  {Nagai}(1986)}]{ProgTheorPhys.76.1237}%
  \BibitemOpen
  \bibfield  {author} {\bibinfo {author} {\bibfnamefont {J.}~\bibnamefont
  {Hara}}\ and\ \bibinfo {author} {\bibfnamefont {K.}~\bibnamefont {Nagai}},\
  }\href {\doibase 10.1143/PTP.76.1237} {\bibfield  {journal} {\bibinfo
  {journal} {Prog. Theor. Phys.}\ }\textbf {\bibinfo {volume} {76}},\ \bibinfo
  {pages} {1237} (\bibinfo {year} {1986})}\BibitemShut {NoStop}%
\bibitem [{\citenamefont {Kwon}\ \emph {et~al.}(2004)\citenamefont {Kwon},
  \citenamefont {Sengupta},\ and\ \citenamefont
  {Yakovenko}}]{EurPhysJB.37.349}%
  \BibitemOpen
  \bibfield  {author} {\bibinfo {author} {\bibfnamefont {H.-J.}\ \bibnamefont
  {Kwon}}, \bibinfo {author} {\bibfnamefont {K.}~\bibnamefont {Sengupta}}, \
  and\ \bibinfo {author} {\bibfnamefont {V.~M.}\ \bibnamefont {Yakovenko}},\
  }\href {\doibase 10.1140/epjb/e2004-00066-4} {\bibfield  {journal} {\bibinfo
  {journal} {Eur. Phys. J. B}\ }\textbf {\bibinfo {volume} {37}},\ \bibinfo
  {pages} {349} (\bibinfo {year} {2004})}\BibitemShut {NoStop}%
\bibitem [{\citenamefont {Kashiwaya}\ and\ \citenamefont
  {Tanaka}(2000)}]{RepProgPhys.63.1641}%
  \BibitemOpen
  \bibfield  {author} {\bibinfo {author} {\bibfnamefont {S.}~\bibnamefont
  {Kashiwaya}}\ and\ \bibinfo {author} {\bibfnamefont {Y.}~\bibnamefont
  {Tanaka}},\ }\href {http://stacks.iop.org/0034-4885/63/i=10/a=202} {\bibfield
   {journal} {\bibinfo  {journal} {Rep. Prog. Phys.}\ }\textbf {\bibinfo
  {volume} {63}},\ \bibinfo {pages} {1641} (\bibinfo {year}
  {2000})}\BibitemShut {NoStop}%
\bibitem [{\citenamefont {Sato}\ \emph {et~al.}(2011)\citenamefont {Sato},
  \citenamefont {Tanaka}, \citenamefont {Yada},\ and\ \citenamefont
  {Yokoyama}}]{PhysRevB.83.224511}%
  \BibitemOpen
  \bibfield  {author} {\bibinfo {author} {\bibfnamefont {M.}~\bibnamefont
  {Sato}}, \bibinfo {author} {\bibfnamefont {Y.}~\bibnamefont {Tanaka}},
  \bibinfo {author} {\bibfnamefont {K.}~\bibnamefont {Yada}}, \ and\ \bibinfo
  {author} {\bibfnamefont {T.}~\bibnamefont {Yokoyama}},\ }\href {\doibase
  10.1103/PhysRevB.83.224511} {\bibfield  {journal} {\bibinfo  {journal} {Phys.
  Rev. B}\ }\textbf {\bibinfo {volume} {83}},\ \bibinfo {pages} {224511}
  (\bibinfo {year} {2011})}\BibitemShut {NoStop}%
\bibitem [{\citenamefont {You}\ \emph {et~al.}(2013)\citenamefont {You},
  \citenamefont {Oh},\ and\ \citenamefont {Vedral}}]{PhysRevB.87.054501}%
  \BibitemOpen
  \bibfield  {author} {\bibinfo {author} {\bibfnamefont {J.}~\bibnamefont
  {You}}, \bibinfo {author} {\bibfnamefont {C.~H.}\ \bibnamefont {Oh}}, \ and\
  \bibinfo {author} {\bibfnamefont {V.}~\bibnamefont {Vedral}},\ }\href
  {\doibase 10.1103/PhysRevB.87.054501} {\bibfield  {journal} {\bibinfo
  {journal} {Phys. Rev. B}\ }\textbf {\bibinfo {volume} {87}},\ \bibinfo
  {pages} {054501} (\bibinfo {year} {2013})}\BibitemShut {NoStop}%
\bibitem [{\citenamefont {Volovik}(1999)}]{JETPLett.70.609}%
  \BibitemOpen
  \bibfield  {author} {\bibinfo {author} {\bibfnamefont {G.~E.}\ \bibnamefont
  {Volovik}},\ }\href {\doibase 10.1134/1.568223} {\bibfield  {journal}
  {\bibinfo  {journal} {JETP Letters}\ }\textbf {\bibinfo {volume} {70}},\
  \bibinfo {pages} {609} (\bibinfo {year} {1999})}\BibitemShut {NoStop}%
\bibitem [{\citenamefont {Furusaki}\ \emph {et~al.}(2001)\citenamefont
  {Furusaki}, \citenamefont {Matsumoto},\ and\ \citenamefont
  {Sigrist}}]{PhysRevB.64.054514}%
  \BibitemOpen
  \bibfield  {author} {\bibinfo {author} {\bibfnamefont {A.}~\bibnamefont
  {Furusaki}}, \bibinfo {author} {\bibfnamefont {M.}~\bibnamefont {Matsumoto}},
  \ and\ \bibinfo {author} {\bibfnamefont {M.}~\bibnamefont {Sigrist}},\ }\href
  {\doibase 10.1103/PhysRevB.64.054514} {\bibfield  {journal} {\bibinfo
  {journal} {Phys. Rev. B}\ }\textbf {\bibinfo {volume} {64}},\ \bibinfo
  {pages} {054514} (\bibinfo {year} {2001})}\BibitemShut {NoStop}%
\bibitem [{\citenamefont {Asahi}\ and\ \citenamefont
  {Nagaosa}(2012)}]{PhysRevB.86.100504}%
  \BibitemOpen
  \bibfield  {author} {\bibinfo {author} {\bibfnamefont {D.}~\bibnamefont
  {Asahi}}\ and\ \bibinfo {author} {\bibfnamefont {N.}~\bibnamefont
  {Nagaosa}},\ }\href {\doibase 10.1103/PhysRevB.86.100504} {\bibfield
  {journal} {\bibinfo  {journal} {Phys. Rev. B}\ }\textbf {\bibinfo {volume}
  {86}},\ \bibinfo {pages} {100504} (\bibinfo {year} {2012})}\BibitemShut
  {NoStop}%
\bibitem [{\citenamefont {Crommie}\ \emph {et~al.}(1993)\citenamefont
  {Crommie}, \citenamefont {Lutz},\ and\ \citenamefont
  {Eigler}}]{Science.262.218}%
  \BibitemOpen
  \bibfield  {author} {\bibinfo {author} {\bibfnamefont {M.~F.}\ \bibnamefont
  {Crommie}}, \bibinfo {author} {\bibfnamefont {C.~P.}\ \bibnamefont {Lutz}}, \
  and\ \bibinfo {author} {\bibfnamefont {D.~M.}\ \bibnamefont {Eigler}},\
  }\href {\doibase 10.1126/science.262.5131.218} {\bibfield  {journal}
  {\bibinfo  {journal} {Science}\ }\textbf {\bibinfo {volume} {262}},\ \bibinfo
  {pages} {218} (\bibinfo {year} {1993})}\BibitemShut {NoStop}%
\bibitem [{\citenamefont {F\"olsch}\ \emph {et~al.}(2004)\citenamefont
  {F\"olsch}, \citenamefont {Hyldgaard}, \citenamefont {Koch},\ and\
  \citenamefont {Ploog}}]{PhysRevLett.92.056803}%
  \BibitemOpen
  \bibfield  {author} {\bibinfo {author} {\bibfnamefont {S.}~\bibnamefont
  {F\"olsch}}, \bibinfo {author} {\bibfnamefont {P.}~\bibnamefont {Hyldgaard}},
  \bibinfo {author} {\bibfnamefont {R.}~\bibnamefont {Koch}}, \ and\ \bibinfo
  {author} {\bibfnamefont {K.~H.}\ \bibnamefont {Ploog}},\ }\href {\doibase
  10.1103/PhysRevLett.92.056803} {\bibfield  {journal} {\bibinfo  {journal}
  {Phys. Rev. Lett.}\ }\textbf {\bibinfo {volume} {92}},\ \bibinfo {pages}
  {056803} (\bibinfo {year} {2004})}\BibitemShut {NoStop}%
\bibitem [{\citenamefont {Hirjibehedin}\ \emph {et~al.}(2006)\citenamefont
  {Hirjibehedin}, \citenamefont {Lutz},\ and\ \citenamefont
  {Heinrich}}]{Science.312.1021}%
  \BibitemOpen
  \bibfield  {author} {\bibinfo {author} {\bibfnamefont {C.~F.}\ \bibnamefont
  {Hirjibehedin}}, \bibinfo {author} {\bibfnamefont {C.~P.}\ \bibnamefont
  {Lutz}}, \ and\ \bibinfo {author} {\bibfnamefont {A.~J.}\ \bibnamefont
  {Heinrich}},\ }\href {\doibase 10.1126/science.1125398} {\bibfield  {journal}
  {\bibinfo  {journal} {Science}\ }\textbf {\bibinfo {volume} {312}},\ \bibinfo
  {pages} {1021} (\bibinfo {year} {2006})}\BibitemShut {NoStop}%
\bibitem [{\citenamefont {Loth}\ \emph {et~al.}(2012)\citenamefont {Loth},
  \citenamefont {Baumann}, \citenamefont {Lutz}, \citenamefont {Eigler},\ and\
  \citenamefont {Heinrich}}]{Science.335.196}%
  \BibitemOpen
  \bibfield  {author} {\bibinfo {author} {\bibfnamefont {S.}~\bibnamefont
  {Loth}}, \bibinfo {author} {\bibfnamefont {S.}~\bibnamefont {Baumann}},
  \bibinfo {author} {\bibfnamefont {C.~P.}\ \bibnamefont {Lutz}}, \bibinfo
  {author} {\bibfnamefont {D.~M.}\ \bibnamefont {Eigler}}, \ and\ \bibinfo
  {author} {\bibfnamefont {A.~J.}\ \bibnamefont {Heinrich}},\ }\href {\doibase
  10.1126/science.1214131} {\bibfield  {journal} {\bibinfo  {journal}
  {Science}\ }\textbf {\bibinfo {volume} {335}},\ \bibinfo {pages} {196}
  (\bibinfo {year} {2012})}\BibitemShut {NoStop}%
\bibitem [{\citenamefont {Okuno}\ \emph {et~al.}(1999)\citenamefont {Okuno},
  \citenamefont {Matsumoto},\ and\ \citenamefont {Sigrist}}]{JPSJ.68.3054}%
  \BibitemOpen
  \bibfield  {author} {\bibinfo {author} {\bibfnamefont {Y.}~\bibnamefont
  {Okuno}}, \bibinfo {author} {\bibfnamefont {M.}~\bibnamefont {Matsumoto}}, \
  and\ \bibinfo {author} {\bibfnamefont {M.}~\bibnamefont {Sigrist}},\ }\href
  {\doibase 10.1143/JPSJ.68.3054} {\bibfield  {journal} {\bibinfo  {journal}
  {J. Phys. Soc. Jpn.}\ }\textbf {\bibinfo {volume} {68}},\ \bibinfo {pages}
  {3054} (\bibinfo {year} {1999})}\BibitemShut {NoStop}%
\end{thebibliography}
\end{document}